\documentclass[reprint,amsmath,amssymb,prc,superscriptaddress,longbibliography]{revtex4-1}

\usepackage{graphicx}
\usepackage[utf8]{inputenc}
\usepackage{color}
\definecolor{goodgreen}{rgb}{0.1,0.5,0}
\definecolor{goodred}{rgb}{0.7,0,0}
\usepackage[colorlinks,urlcolor=goodgreen,citecolor=blue,linkcolor=goodred]{hyperref}
\usepackage{upgreek}

\usepackage{layouts}

\usepackage{newtxtext,newtxmath}

\renewcommand{\text}[1]{\ensuremath{\mathrm{#1}}}
\newcommand{\un}[1]{\ensuremath{\,\textrm{#1}}}

\newcommand{\fdrive}{\ensuremath{f_\text{d}}}

\newcommand{\fprobe}{\ensuremath{f_\text{p}}}
\newcommand{\wcav}{\ensuremath{\omega_\text{c}}}
\newcommand{\fcav}{\ensuremath{f_\text{c}}}
\newcommand{\wmech}{\ensuremath{\omega_\text{m}}}
\newcommand{\fmech}{\ensuremath{f_\text{m}}}

\newcommand{\Gacav}{\ensuremath{\Gamma_\text{c}}}
\newcommand{\kcav}{\Gacav}

\newcommand{\Qcav}{\ensuremath{Q_\text{c}}}
\newcommand{\Ecav}{\ensuremath{E_\text{cav}}}
\newcommand{\Gamech}{\ensuremath{\Gamma_\text{m}}}
\newcommand{\Gameff}{\ensuremath{\Gamma_\text{eff}}}
\newcommand{\kmech}{\Gamech}

\newcommand{\Qmech}{\ensuremath{Q_\text{m}}}
\newcommand{\Vg}{\ensuremath{V_\text{g}}}
\newcommand{\Cg}{\ensuremath{C_\text{g}}}
\newcommand{\Qg}{\ensuremath{Q_\text{g}}}
\newcommand{\Cx}{\ensuremath{C_\text{x}}}
\newcommand{\Ce}{\ensuremath{C_\text{e}}}
\newcommand{\Cq}{\ensuremath{C_\text{q}}}
\newcommand{\RRx}{\ensuremath{R_\text{x}}}
\newcommand{\RRe}{\ensuremath{R_\text{e}}}

\newcommand{\Qdot}{\ensuremath{Q_\text{dot}}}

\newcommand{\Vsd}{\ensuremath{V_\text{sd}}}

\newcommand{\vg}{\Vg}
\newcommand{\Ccav}{\ensuremath{C_{\text{c}}}}

\newcommand{\xzpf}{\ensuremath{x_{\text{zpf}}}}
\newcommand{\avn}{\ensuremath{\left<N\right>}}
\newcommand{\ncav}{\ensuremath{n_\text{c}}}

\newcommand{\vsd}{\ensuremath{V_{\text{sd}}}}

\newcommand{\vgac}{\ensuremath{V_\text{g}^\text{ac}}}

\newcommand{\ti}[1]{_\text{#1}}
\newcommand{\CRLC}{\ensuremath{C_\text{RLC}}}
\newcommand{\RRLC}{\ensuremath{R_\text{RLC}}}
\newcommand{\LRLC}{\ensuremath{L_\text{RLC}}}

\begin{document}

\title{Optomechanical coupling and damping of a carbon nanotube quantum dot} 

\author{N. Hüttner}
\author{S. Blien}
\author{P. Steger}
\altaffiliation[Current address: ]{Department of Physics, Lancaster University, 
Lancaster LA1 4YB, 
United Kingdom}
\author{A. N. Loh}
\author{R. Graaf}
\affiliation{Institute for Experimental and Applied Physics, 
University of Regensburg, Universitätsstr.\ 31, 93053 Regensburg, Germany}
\author{A. K. Hüttel}
\email{andreas.huettel@ur.de}
\affiliation{Institute for Experimental and Applied Physics, 
University of Regensburg, Universitätsstr.\ 31, 93053 Regensburg, Germany}
\affiliation{Department of Applied Physics, Aalto University, 
Puumiehenkuja 2, 02150 Espoo, Finland}

\begin{abstract}
Carbon nanotubes are excellent nano-electromechanical systems, combining high
resonance frequency, low mass, and large zero-point motion. At cryogenic
temperatures they display high mechanical quality factors. Equally they are
outstanding single electron devices with well-known quantum levels and have
been proposed for the implementation of charge or spin qubits. The integration 
of these devices into microwave optomechanical circuits is however hindered by a
mismatch of scales, between typical microwave wavelengths, nanotube segment
lengths, and nanotube deflections. As experimentally demonstrated recently in
[Blien {\it et al.}, Nat. Comm. {\bf 11}, 1363 (2020)], coupling enhancement 
via the quantum capacitance allows to circumvent this restriction. Here we 
extend the discussion of this experiment. We present the subsystems of the 
device and their interactions in detail. An alternative approach to the 
optomechanical coupling is presented, allowing to estimate the mechanical zero 
point motion scale. Further, the mechanical damping is discussed, hinting at 
hitherto unknown interaction mechanisms.
\end{abstract}

\maketitle 

\tableofcontents

\section{Introduction}

Optomechanics \cite{rmp-aspelmeyer-2014} and its manifold branches allow the 
characterization and manipulation of both macroscopic and nanoscale mechanical 
systems. By now readily available techniques include, e.\ g., ground state 
cooling \cite{nature-chan-2011, nature-teufel-2011} and squeezing 
\cite{prx-lecocq-2015} of nanomechanical states, displacement sensing at and 
beyond the standard quantum limit \cite{nnano-teufel-2009}, or on chip optical 
data processing \cite{apr-metcalfe-2014}. Optomechanical techniques and 
formalisms have been applied to a wide range of material systems, from single 
atoms in traps to macroscopic interferometer mirrors \cite{rmp-aspelmeyer-2014}.

Suspended single-wall carbon nanotubes (SW-CNTs) as mechanical resonators have 
been shown to reach high quality factors of up to $5\times 10^6$ \cite{highq,
nnano-moser-2014} in a cryogenic environment. At the same time, they are
excellent quantum dots and clean electronic quantum mechanical model systems, 
and transport spectroscopy at millikelvin temperatures has led to a large 
number of topical publications \cite{rmp-laird-2015, highfield, transparency}. 
The observation of strong coupling between single electron tunneling and the 
motion of the macromolecule \cite{strongcoupling, science-lassagne-2009} has
inititated a further field of research \cite{highqset, nl-hakkinen-2015,
kondocharge}, as has the integration of carbon nanotubes into circuit cavity
quantum electrodynamics experiments \cite{science-viennot-2015,
nature-desjardins-2017}.

Regarding the combination of the two fields, experimental approaches for 
optomechanics with carbon nanotubes at optical / visible frequencies exist 
\cite{apl-stapfner-2013, zhang_cavity_2014, ncomms-tavernarakis-2018, 
nature-barnard-2019}. However, since the photon energy exceeds the typical 
energy range of trapped electronic quantum states at low temperature, excitonic 
states, or even the electronic band gap, they are fundamentally incompatible 
with Coulomb blockade experiments. Consequently, this frequency range needs to 
be excluded from consideration in all experiments where the electronic 
confinement within the nanotube plays a role.

The small dimensions of typical single electron devices prevent effective
integration into microwave optomechanical systems via conventional mechanisms
relying only on radiation pressure \cite{rmp-aspelmeyer-2014,
nphys-regal-2008}. This mismatch of scales, critical for quasi one-dimensional 
objects as compared to, e.g., nanomechanical drum resonators 
\cite{nature-teufel-2011, arxiv-das-2023}, becomes immediately obvious when 
comparing the typical microwave wavelength and thereby resonator size, $\sim 
1\un{cm}$ for $\fcav\sim 5\un{GHz}$, the typical length of a suspended carbon 
nanotube quantum dot $\sim 1\,\mu\text{m}$, and the typical deflection of such a 
suspended nanotube $\sim 1\un{nm}$.

Recently, we have shown that the large variation in quantum capacitance of a
CNT in the Coulomb blockade regime enhances the optomechanical coupling by
several orders of magnitude at suitable choice of a working point 
\cite{optomechanics}. Using optomechanically induced (in)transparency 
\cite{pra-agarwal-2010, science-weis-2010}, a single photon coupling of
up to $g_0 = 2\pi \times 95\un{Hz}$ was measured. Here, we expand upon the data 
evaluation and discussion of \cite{optomechanics} and characterize a wide range 
of interactions in the device already partly presented there --- between the
mechanical resonator, the microwave resonator, and the quantum dot in the 
Coulomb blockade regime. Combining different types of measurements, we provide 
an extended model, which allows us to estimate, e.g., the zero point motion 
amplitude of the carbon nanotube, further discuss consistency of the resulting 
device parameters, and characterize the mechanical damping mechanisms.

\section{Device and measurement setup}

\begin{figure}[t]
\begin{center}
\includegraphics{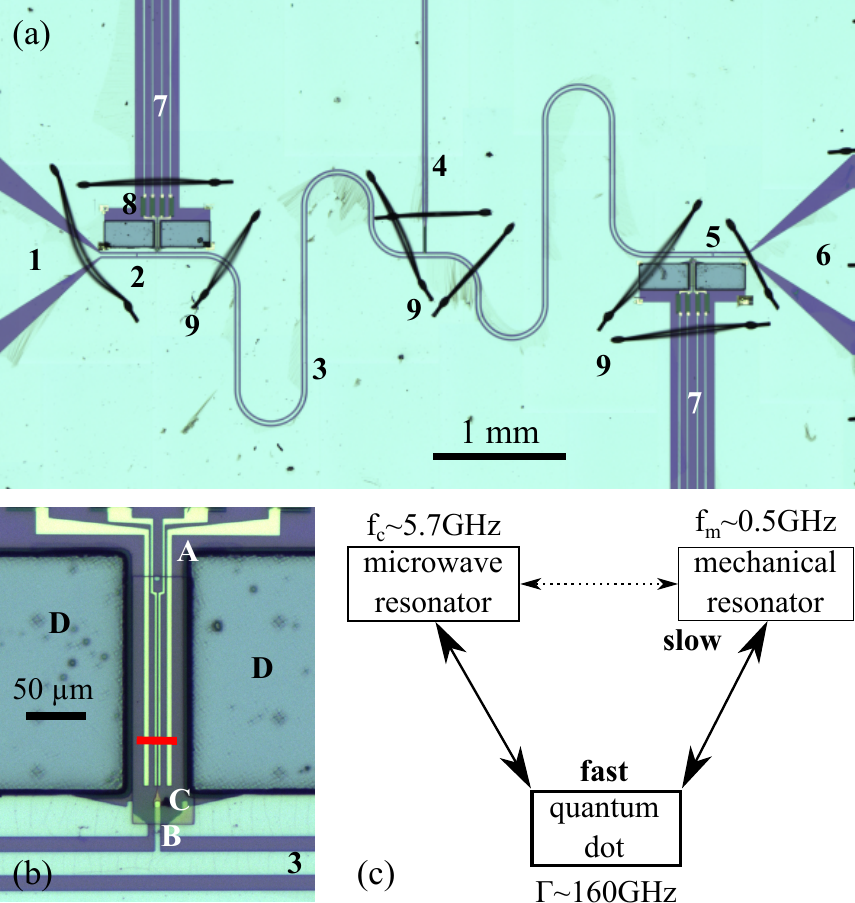}
\end{center}
\caption{(a) Optical microscope overview image of the device, combining a
coplanar waveguide resonator and two carbon nanotube deposition areas (only 
one of which was used in the measurement).  1: GHz input port, 2: input coupling
capacitor, 3: coplanar waveguide resonator, 4: dc (gate voltage) connection, 5: 
output coupling capacitor, 6: GHz output port, 7: dc (source, drain, and 
cutting electrode) connections, 8: meander inductance filters, 9: bond wires 
for potential equilibration. (b) Detail image of the nanotube deposition area. 
A: Source-, drain-, and (2x) outer cutting electrodes, B: gate finger connected 
to the coplanar resonator, C: gate isolation (transparent, cross-linked PMMA), 
D: deep-etched ($\sim 10\un{\textmu m}$) region to allow for fork deposition of 
the nanotubes. The red line corresponds to the cut sketched in 
Fig.~\ref{figCNTDot}(a) and thus also to one possible location of a carbon 
nanotube. (c) Scheme detailing the interactions between nanotube and 
microwave field and the relevant frequencies. Microscope images adapted from 
\cite{optomechanics}, Fig. S1.
\label{figSampleMind}
}
\end{figure}
 
Our device, shown in Fig.~\ref{figSampleMind}(a), combines a superconducting
coplanar microwave cavity with a suspended CNT quantum dot, this way acting as
optomechanical hybrid structure. Coupling between the two subsystems is 
mediated via a gate electrode. This gate electrode, buried below the nanotube,
is connected to the center conductor of the microwave resonator close to its
input coupling capacitance, i.e., at one of the electric field and voltage
antinodes.

Carbon nanotube and coplanar waveguide resonator form separate circuits,
coupling to each other only capacitively. The CNT displays Coulomb blockade
oscillations of conductance as function of gate voltage, but also acts as a 
high-$Q$ mechanical oscillator. Initially, the CNT is characterized via 
standard low-frequency quantum dot transport spectroscopy \cite{kouwenhoven, 
rmp-laird-2015}, and the resonator via a GHz transmission measurement.

\subsection{Niobium coplanar resonator}

The microwave-optical subsystem of our device is given by a coplanar
half-wavelength resonator \cite{book-pozar, book-simons, ieee-gevorgian-1995}.
On a high-resistivity ($>10\,\text{k}\Omega\text{cm}$) float-zone silicon
substrate with a 500\,nm thermally grown surface oxide, a uniform niobium layer
is sputter-deposited. Using standard optical lithography followed by reactive
ion etching, an impedance-matched coplanar waveguide (CPW) with in/out couple
bond pads at both ends for transmission measurement is defined, see
Fig.~\ref{figSampleMind}(a). Its center conductor {\bf 3} is interrupted by 
gaps twice, see {\bf 2} and {\bf 5} in Fig.~\ref{figSampleMind}(a), forming a 
$\lambda/2$ type resonant cavity for transmission measurement. The gap width 
determines the coupling capacitances $C_\text{in}$ and $C_\text{out}$ across 
which the resonator is driven and read out; the distance between the gaps 
defines the resonator length $\ell=10.5\un{mm}$ and with it on first order the 
fundamental resonance frequency.

Close to the coupling capacitances, i.e., at the voltage antinodes, a stub-like
extension of the center conductor connects to the gate electrode for the carbon
nanotube, see {\bf B} in Fig.~\ref{figSampleMind}(b) and the discussion below.
Additionally, at its midpoint, i.e., the voltage node of the fundamental
electromagnetic resonance mode, the center conductor is connected to a dc feed, 
{\bf 4} in Fig.~\ref{figSampleMind}(a), for the application of a gate bias 
\cite{nature-petersson-2012}. This dc feed, similar to the connections to the 
carbon nanotube discussed below, contains a thin, resistive gold meander with 
an approximate length of 3\,mm acting as radio-frequency block.

\begin{figure}[t]
\begin{center}
\includegraphics{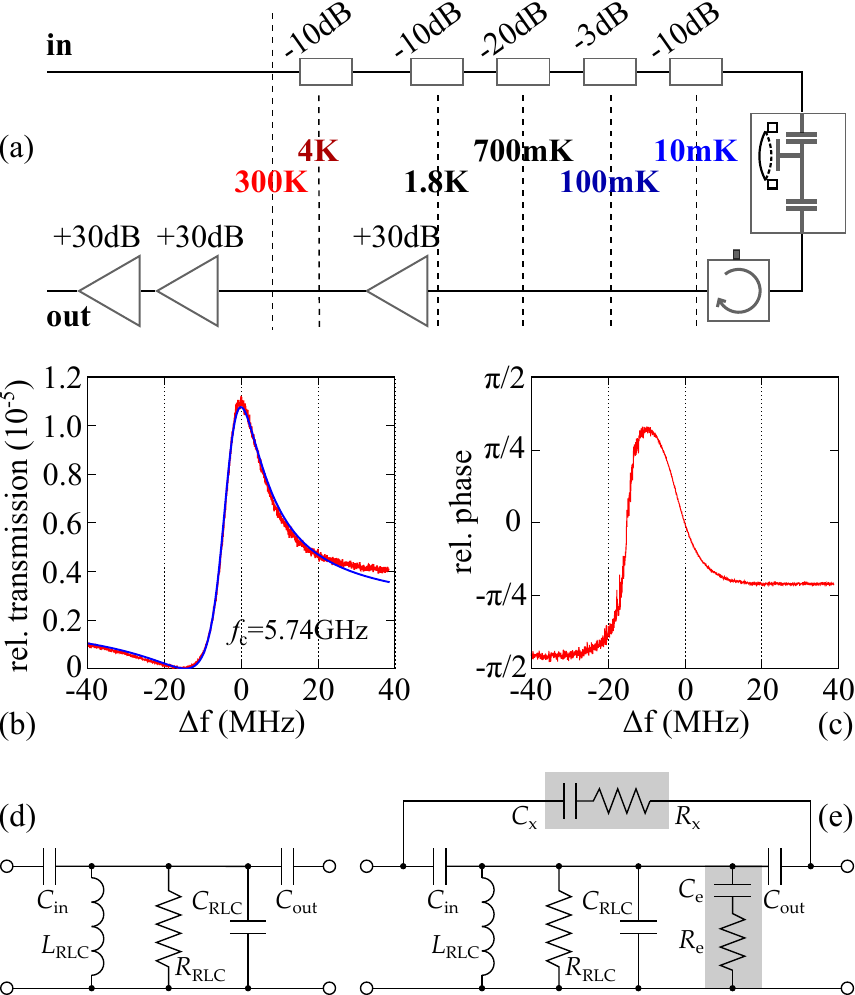}
\end{center}
\caption{(a) Schematic GHz transmission measurement wiring of the dilution 
refrigerator, with attenuators for thermalization at each temperature stage on 
the input side, a cryogenic circulator (QuinStar QCY‐060400CM00), a cryogenic  
amplifier (CalTech CITCRYO1-12A), and two room-temperature amplifiers at the 
output side. (b,c) Transmission amplitude (b, data identical to 
\cite{optomechanics}, Fig. S9) and phase angle (c) of the 
measured coplanar resonator near resonance. The maximum transmission is
$-49.6\un{dB}$. The blue line is a fit using the
phenomenological Fano model of Eq.~(\ref{eq-s21fano}). (d) Generic replacement 
RLC circuit of a coupled waveguide resonator measured in transmission 
\cite{jap-goeppl-2008}. (e) Extended replacement RLC circuit taking into 
account the CNT circuitry (via $C\ti{e}$ and $R\ti{e}$) and cross-talk through 
the sample space (via $C\ti{x}$ and $R\ti{x}$). \label{figCoplanarRes}}
\end{figure}
Figure~\ref{figCoplanarRes}(a) schematically shows the cryogenic measurement 
setup, Fig.~\ref{figCoplanarRes}(b) an example transmission measurement $\left| 
S\ti{21} \right|^2$ of the resonator at base temperature $T \simeq 10\un{mK}$ 
of the dilution refrigerator, and Fig.~\ref{figCoplanarRes}(c) the 
corresponding transmission phase. The measured value includes cable damping of 
approximately $-8\un{dB}$, the attenuators of $-53 \un{dB}$ distributed over
the stages of the dilution refrigerator for input cable thermalization, a low 
temperature HEMT amplifier at the 1.8\,K stage with amplification of 
$30\un{dB}$, and a room temperature amplifier chain with a total amplification 
of approximately $60\un{dB}$.

\begin{table}
\begin{center}
\begin{tabular}{|p{3.5cm} |c|c|c|}
	\hline 
	\textbf{Microwave cavity} & & & \\
	Cavity resonance frequency & $^{(1)}$ &  \fcav & $5.74005\un{GHz}$ \\
	Cavity line width & $^{(1)}$ & $\kcav$ & $2\pi\cdot 11.6\,\text{MHz}$  \\
	Cavity quality factor & $2\pi\fcav/\kcav$ & $Q_{\text{c}}$ & $497$ \\
	Cavity total capacitance & $^{(2)}$ & \Ccav & $1750\un{fF}$ \\
	Replacement capacitance & $\Ccav/2$ & \CRLC & $875\un{fF}$ \\
	Replacement inductance & $1/(4\pi^2\fcav^2\CRLC)$ & \LRLC & $879\un{pH}$ 
\\
	\hline
\end{tabular}
\end{center}
\caption{
Overview of the microwave cavity parameters. $^{(1)}$Obtained from a 
fit using Eqn.~(\ref{eq-s21fano}). $^{(2)}$Calculated CPW capacitance from 
lithographic geometry of the waveguide and substrate material properties, and
neglecting the much smaller $C_\text{in}$, $C_\text{out}$, and \Cg.
\label{par-cpw}}
\end{table} 

Any coplanar waveguide resonator can be translated into a lumped element RLC
replacement circuit with identical resonant frequency $\wcav = 1 / \sqrt{
L_{RLC} C_{RLC} }$ \cite{jap-goeppl-2008}. Fig.~\ref{figCoplanarRes}(d) displays
the simplest such variant for a $\lambda/2$ resonator measured in transmission
\cite{jap-goeppl-2008}. The relationship $\CRLC=\Ccav/2$ (for the fundamental 
resonance only) effectively expresses that fields are not distributed equally 
along the CPW cavity, with an electric field node of this mode at its center. 

Our measurement displays a clear Fano shape instead of the Lorentzian naively 
expected from the circuit of Fig.~\ref{figCoplanarRes}(d), indicating the 
presence of additional non-resonant transmission channels parallel to the 
coplanar resonator. In terms of a circuit model, such a Fano shape can be taken 
into account by introducing a parallel channel \cite{arxiv-hornibrook-2012}, 
see Fig.~\ref{figCoplanarRes}(e) and in particular \Cx\ and \RRx\ there. In 
addition, the figure introduces the impact of a coupled carbon nanotube and its 
electrodes, via \Ce\ and \RRe. 

Comparisons have however shown that it makes no significant difference for our 
evaluation whether we calculate the $S_{21}$ parameter for 
Fig.~\ref{figCoplanarRes}(e) analytically or work with a conceptually much 
simpler Fano model that absorbs \Ce\ and \RRe\ into \CRLC\ and \RRLC\ and takes 
\Cx\ and \RRx\ into account via a complex constant offset of $S_{21}$. In this 
model one obtains for the transmission \cite{apl-khalil-2012, jap-petersan-1998}
\begin{equation}\label{eq-s21fano}
S_{21} =
 A \left(
   \frac{1}{1+2 i Q\ti{c}\; (f-f_0)/f_0} + r e^{i\theta}
 \right), 
\end{equation}
where $r$ and $\theta$ describe transmission amplitude and phase of the 
parallel, parasitic channel. Using a fit of Eq.~(\ref{eq-s21fano}), we obtain 
from the measurement a resonance frequency $\fcav = 5.74 \un{GHz}$ and a 
resonance width of $\Gacav = 2\pi\times 11.6 \, \text{MHz}$, corresponding to a 
quality factor $Q\ti{c} = 497$. Table~\ref{par-cpw} collects these device 
parameters as an overview.

In comparison with similar experiments from literature \cite{nphys-regal-2008,
nnano-singh-2014}, the observed quality factor of our device is surprisingly
low. Given that we have already fabricated coplanar waveguide resonators with 
intrinsic quality factors of $Q\ti{i} \simeq 2\times 10^5$ \cite{toscres}, the 
limitation is likely not given by resonator or substrate material or the 
resonator patterning process per se. Two factors contribute here. On the one
hand, the multiple lithographic steps required for device fabrication lead to
an increased chance of defects and contamination. Examples can be seen in
Fig.~\ref{figSampleMind}(a) near the center of the coplanar waveguide 
resonator, with veil-like structures from fluorinated resist residues. Testing
of multiple resonator devices has shown that such structures on top of the
center conductor lead to a significant decrease of the quality factor.

On the other hand, the dc electrodes of the carbon nanotube deposition areas
couple out part of the GHz signal from the coplanar resonator. While the gold
meanders in the dc connections are intended as inductive high-frequency blocks,
they are also resistive, leading to signal dissipation. Future device design
shall replace them with reflective low-pass filters \cite{apl-hao-2014} to
avoid signal loss near the cavity resonance.

\begin{figure}[t]
\begin{center}
\includegraphics{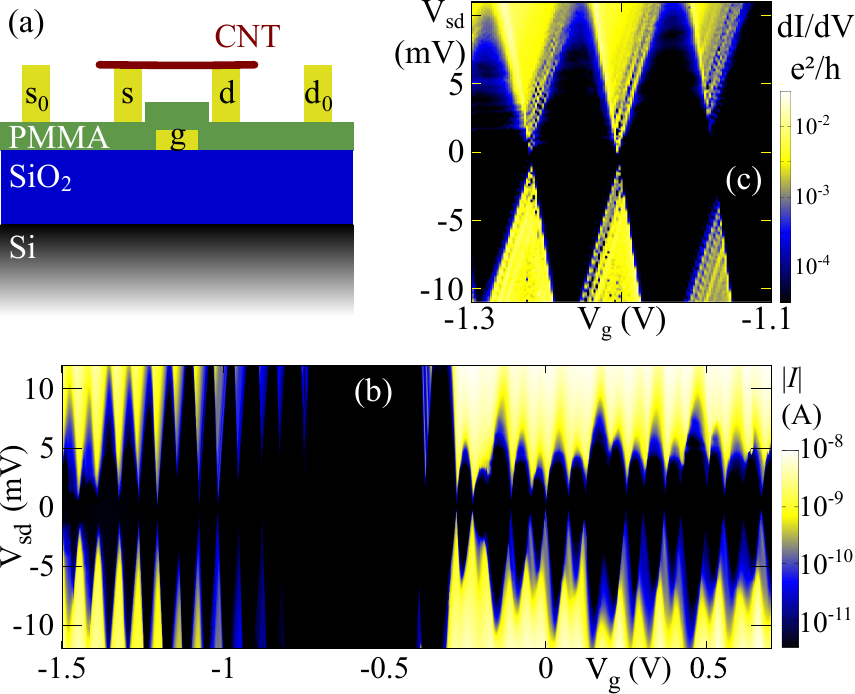}
\end{center}
\caption{(a) Schematic side view of the carbon nanotube lying across contacts 
and gate and forming a typical quantum dot device. A possible location for this 
trace cut is given in Fig.~\ref{figSampleMind}(b) with a red line. (b) Overview 
plot of the current $\left|I_\text{sd}(\vg,\vsd)\right|$ as function of gate 
voltage \vg\ and bias voltage \vsd, showing Coulomb blockade oscillations on 
both sides of the presumed electronic band gap of the nanotube (larger range of 
the data of \cite{optomechanics}, Fig. 2(c)). (c) Detail from (b); plot of the 
numerically obtained differential conductance $\text{d} I_\text{sd} / \text{d} 
\vsd(\vg, \vsd)$ in the parameter region later characterized in GHz 
measurements, again displaying the typical regions of Coulomb blockade and of 
single electron tunneling. 
\label{figCNTDot}
}
\end{figure}

\subsection{Carbon nanotube quantum dot}

The device includes two regions where a carbon nanotube can be deposited onto
contacts next to the coplanar waveguide resonator, see
Fig.~\ref{figSampleMind}(b). For the measurements presented here, only one of 
these was used. The detailed carbon nanotube growth and transfer procedure, 
adapted from works of other research groups in the field, has already been 
discussed in detail elsewhere \cite{nl-wu:1032, forktransfer, optomechanics}. 
After deposition, the nanotube freely crosses a trench of width $L = 
1\,\mu\text{m}$ between two gold electrodes acting as source and drain. A 
finger gate at the bottom of the trench, below an isolating PMMA layer, is 
connected to the coplanar waveguide resonator for coupling, and can also be 
used to apply a dc gate voltage, see Fig.~\ref{figCNTDot}(a).

When varying the applied gate voltage, we observe the typical Coulomb blockade 
oscillations of a quantum dot \cite{kouwenhoven, nature-tans-1997,
rmp-laird-2015}, see Fig.~\ref{figCNTDot}(b). In this overview plot of the dc 
current, $\left|I(\vg,\vsd)\right|$, an apparent electronic band gap around 
$\vg=-0.6\un{V}$ is flanked on both sides by Coulomb blockade oscillations. A 
detail measurement of the differential conductance in the parameter region 
later used for the optomechanical measurements, Fig.~\ref{figCNTDot}(c), 
displays multiple differential conductance lines in single electron tunneling, 
possibly related to longitudinal vibration \cite{prb-braig-2003, prl-koch-2005, 
prl-sapmaz-2006, cocoset, franckcondon}, electronic excitations, or trap states 
in the contacts. No clear fourfold shell pattern of the Coulomb oscillations can 
be observed, possibly due to small-scale defects or disorder of the CNT.

\begin{table}
\begin{center}
\begin{tabular}{|p{4cm} |c|c|c|c|c|}
	\hline 
	\textbf{Nanotube quantum dot} & & & \\
	Gate capacitance & $^{(1)}$ & \Cg & 2.6\,aF \\
	Total capacitance & $^{(1)}$ & $C_\Sigma$ & 9.8\,aF \\
	Gate lever arm & $\Cg/C_\Sigma$ & $\alpha$ & 0.27 \\
	Total tunnel rate & $^{(1)}$ & $\Gamma$ & 160\,GHz \\
	Effective electronic length & & $\ell_\text{eff}$ & $140\un{nm}$ \\
	\hline
	\textbf{Nanotube mechan. resonator} & & & \\
	Mode 1 curvature & $^{(2)}$ & $a_1$ & $-20.42\un{kHz/V}^2$ \\
	Mode 1 center voltage & $^{(2)}$& $V_{g0,1}$ & $-2.64234\un{V}$ \\
	Mode 1 center frequency & $^{(2)}$& $f_{0,1}$ & $502.592\un{MHz}$ \\
	Mode 2 curvature & $^{(2)}$& $a_2$ & $6.41\un{kHz/V}^2$ \\
	Mode 2 center voltage & $^{(2)}$& $V_{g0,2}$ & $7.47759\un{V}$ \\
	Mode 2 center frequency & $^{(2)}$& $f_{0,2}$ & $500.537\un{MHz}$ \\
	Normal mode splitting 1-2 & $^{(2)}$& $\Delta f_\text{min}$ & 
$450\un{kHz}$ \\
	Suspended length of nanotube & $^{(3)}$ & $\ell$ & $1\,\mu\text{m}$ \\
	Radius of nanotube & $^{(4)}$ & $r$ & $2\un{nm}$ \\
	Effective mass & $^{(4)}$ & $m$ & $4.8\cdot 10^{-21} \un{kg}$  \\
	Imprinted tension & $^{(5)}$ & $T_0$ & $4.8\un{nN}$ \\
	Mech. line width & & $\kmech$ & $\lesssim 2\pi\cdot 50\un{kHz} $ \\
	Quality factor & $2\pi\fmech/\kmech$ & $Q_\text{m}$ & $\gtrsim 10^4$ \\
	\hline
\end{tabular}
\end{center}
\caption{
Overview of the carbon nanotube parameters. $^{(1)}$From Coulomb blockade 
characterization near $\vg=-1.2\un{V}$, see Fig.~\ref{figCNTDot}(b).
$^{(2)}$From the coupled classical harmonic oscillator fit using
Eq.~(\ref{eq:mechfit}); see also Fig.~\ref{figMechanics}(a).
$^{(3)}$Lithographic distance of the contact electrodes. $^{(4)}$Estimated,
typical values. $^{(5)}$Calculated via Eq.~(\ref{eq:tension}).
\label{par-cnt}
}
\end{table}

In the parameter region discussed in detail below we obtain via evaluation of
the Coulomb blockade data of Fig.~\ref{figCNTDot}(b,c) \cite{kouwenhoven} for
the quantum dot capacitances of $\Cg=2.6\un{aF}$ and $C_\Sigma=9.8\un{aF}$, and
with these the gate conversion factor $\alpha=0.27$, see also
Table~\ref{par-cnt}. In addition, we can estimate the total tunnel rate of
quantum dot--lead coupling from the zero-bias conductance peak broadening. A
value of $\Gamma=160\un{GHz}$, corresponding to $0.69\un{meV}$, is consistent
both with conductance and (as discussed later) optomechanical coupling
\cite{optomechanics}. This makes electronic tunneling the fastest relevant time 
scale in our coupled system, cf. Fig.~\ref{figSampleMind}(c), clearly exceeding 
the cavity and mechanical resonance frequencies.

An estimation of the gate capacitance via a simple wire-over-plane
model \cite{optomechanics, apl-wunnicke-2006}, using an averaged relative
dielectric constant $\epsilon_r=2$ and a gate distance $d=450\un{nm}$, coincides
with the gate capacitance from Coulomb blockade for a reduced length
$\ell_\text{eff} = 140\un{nm}$ of the nanotube; we call this the {\em effective
electronic length} of our carbon nanotube quantum dot. This models, e.g., the
reduction caused by depletion regions in pn-barriers, or more generally corrects
for geometry deviations.

\begin{figure}[t]
\begin{center}
\includegraphics{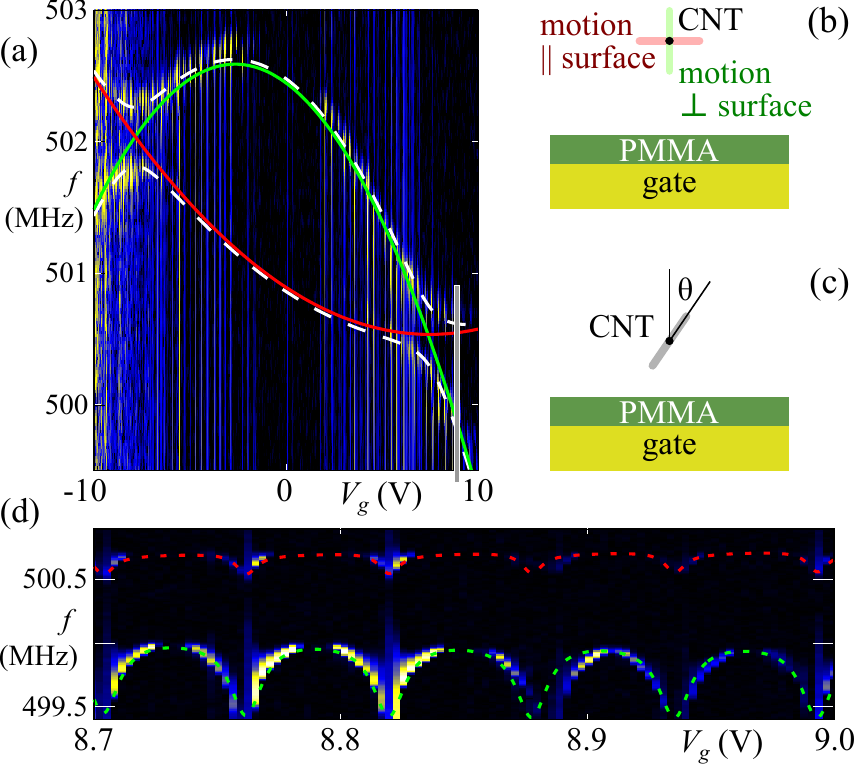}
\end{center}
\caption{
(a) Large-scale plot of the mechanical resonance detection result, combined 
with a fit of two coupled vibration modes of parabolic gate voltage dependence. 
An amplitude-modulated ($f_\text{am}=72\un{Hz}$) driving signal at frequency 
$f$ is applied via a contact-free antenna \cite{highq, strongcoupling, 
highqset, magdamping}; the plot shows the resulting modulation of the rectified 
current via the lock-in signal $\text{d}I / \text{d}P_\text{rf}$. The dashed 
lines are a fit using Eq.~\ref{eq:mechfit}; see Table~\ref{par-cnt} for the fit 
parameters. The red and green solid lines show the two corresponding 
vibration modes in absence of coupling. Data already shown in 
\cite{optomechanics}, Figs. S6 and S7.
(b) Schematic drawing (view along the carbon nanotube axis) of the two 
transversal vibration modes parallel and perpendicular to the device surface.
(c) Schematic of a transversal vibration mode with arbitrary orientation angle 
$\theta$.
(d) Coulomb oscillations of mechanical resonance frequency 
\cite{science-lassagne-2009, strongcoupling}, in the parameter region marked in 
(a) with a grey rectangle, and using the same measurement scheme as in (a).
The dashed lines indicate a fit with subsequent Coulomb oscillations as 
described in the text. Raw data already shown in \cite{optomechanics}, Fig. S8.
\label{figMechanics}
}
\end{figure}

\subsection{Driven vibrational motion of the nanotube}

The behavior of a carbon nanotube quantum dot as high-$Q$ nanomechanical 
resonator at cryogenic temperatures has been discussed in many recent works 
\cite{highq, strongcoupling, nnano-moser-2014, nl-hakkinen-2015, heliumdamping, 
kondocharge, ncomms-rechnitz-2022}. The dominant external force acting on the 
nanotube as a suspended beam is given by the electrostatic force of the gate 
charge acting on the quantum dot charge; the restoring force stems from the 
tension, either built-in or deflection induced, and the bending rigidity of the 
macromolecule. Overall deflection leads to an elongation, increase of the 
tension, and thereby an increase of its mechanical resonance frequency 
\cite{nature-sazonova-2004, nl-witkamp-2006}. 

The fundamental bending mode resonance frequency of a suspended nanotube scales 
with the segment length $L$ as $1/L^2$; from literature we typically expect it 
in the range of $50\un{MHz} \le f_m \le 100\un{MHz}$ for a $L=1\,\mu\text{m}$ 
long nanotube \cite{nature-sazonova-2004, nl-witkamp-2006, highq, magdamping, 
mechtransport}. The device presented here shows two resonances around $\sim 
500\un{MHz}$, see Fig.~\ref{figMechanics}(a). In the plot, a lock-in amplifier 
is used to amplitude-modulate the applied rf driving signal at $f_\text{am} = 
72\un{Hz}$ and pick up the corresponding modulation of the low-frequency 
current through the nanotube; while slightly less sensitive than the so-called 
frequency modulation technique \cite{small-gouttenoire-2010}, this method 
retains a more natural resonance shape.

A thorough search at lower drive frequencies led to no additional results. In 
combination with the weak gate voltage dependence, this indicates a high 
built-in tension imprinted onto the carbon nanotube during fabrication. Its 
origin likely lies in the transfer of carbon nanotubes into the resonator 
circuit \cite{nl-wu:1032, forktransfer, optomechanics}. The nanotubes are grown 
on a quartz fork and lowered onto the contact electrodes until a finite current 
is measured. Then, they are locally cut by resistive heating with a large 
current through it. The heating only affects the nanotube and its 
immediate surroundings, with the macromolecule slightly melting the gold 
contacts and attaching there; traces of this have been observed in microscope 
images of test structures. The force pulling the nanotube over the contacts
leads to an eventual imprinted tension in the device.
\begin{figure}[t]
\begin{center}
\includegraphics{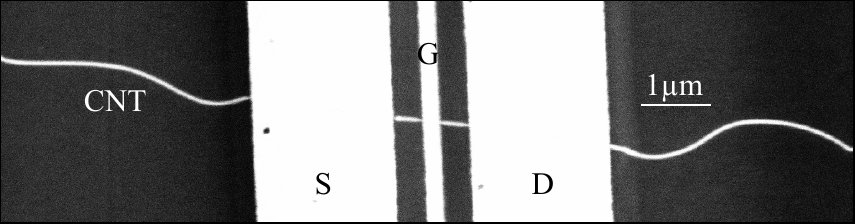}
\end{center}
\caption{Detail SEM image of a different device fabricated later, but having a
similar fabrication procedure and geometry. A carbon nanotube lies across
source, gate, and drain electrode. Between the source and drain electrodes it
is stretched straight; outside them it bends.
\label{figStretch}
}
\end{figure}
This effect is also illustrated by the SEM image of Fig.~\ref{figStretch}. It
shows a different, subsequently produced device, where however similar
fabrication steps and lithographic geometries have been used. Between source (S)
and drain (D) electrode, the visible nanotube (CNT) is stretched straight, while
it is clearly non-tensioned outside these electrodes.

Cooling of the device from room temperature to base temperature of the dilution
refrigerator leads to a thermal contraction of the silicon substrate. 
Even though a negative axial coefficient of expansion for carbon nanotubes has 
been theoretically predicted for a long time \cite{prl-kwon-2004, 
prb-mounet-2005, nmat-balandin-2011}, surprisingly few experiments exist  
\cite{jpcc-chi-2018}. What data there is confirms a negative thermal expansion, 
hinting that the cooling of the nanotube and the expansion coefficient mismatch 
with the substrate should not introduce tension (but rather counteract tension 
or induce buckling).

Our observed modes in the tensioned case correspond in first approximation 
to the fundamental transversal vibration parallel to the device surface and 
towards the gate \cite{apl-kozinsky-2006}, see Fig.~\ref{figMechanics}(b): only 
for the latter, electrodynamic softening of the vibration mode 
\cite{apl-kozinsky-2006, nl-wu-2011, negtuning} contributes to the large-scale 
gate voltage dependence of the resonance frequency, inverting the dispersion at 
low gate voltage. The negative curvature term is absent for motion parallel to 
the chip surface, where only the tension-induced frequency increase is 
observed. Note that this model, treating the nanotube as a two-dimensional 
oscillator, still simplifies away many physically relevant details, from the 
deflection envelope along the nanotube all the way to screw-like motions or 
buckling \cite{ncomms-rechnitz-2022}.

As fit functions in Fig.~\ref{figMechanics}(a), two coupled classical harmonic 
oscillator modes with general parabolic dispersion 
\begin{equation}
f_i(\Vg) = a_i(V_g - 
V_{g0,i})^2 + f_{0,i}
\end{equation}
are chosen ($i=1,2$; solid lines in the figure), leading to fit functions
\begin{equation}\label{eq:mechfit}
 f_\pm(\Vg) = \sqrt{\frac{1}{2}\left( f_1^2 + f_2^2 \right)
 \pm \frac{1}{2} \sqrt{\left( f_1^2 - f_2^2 \right)^2 + 4 W^2}}
\end{equation}
(dashed lines in the figure). Here, $W$ parametrizes the coupling between the
two modes. In the evaluated voltage range, this model
describes the large-scale gate voltage dependence of the resonance frequencies 
very well; the resulting fit parameters can be found in Table~\ref{par-cnt}. 
The coupling of the vibration modes via the tension of the nanotube induces a 
sizeable mode splitting of $0.45\un{MHz}$, similar in magnitude to previous 
observations \cite{prl-eichler-2012}.

With the effective mass $m=4.8\cdot 10^{-21}\un{kg}$, $f_\text{max} \simeq 
501.5\un{MHz}$, and the suspended length $\ell = 1\,\mu\text{m}$, we obtain 
using the relation
\begin{equation}\label{eq:tension}
 f_\text{max}= \frac{1}{2}\sqrt{\frac{T_0}{m\ell}}
\end{equation}
the remarkably large fabrication-imprinted tension (i.e., the axial tension of 
the nanotube in absence of electrostatic forces) $T_0 = 4.8\un{nN}$.

Assuming now that the contributions to the spring constant add up linearly, we 
can use the difference in curvature of the two modes to isolate the
electrostatic softening effects alone. With $a_s = a_1 - a_2$, following 
\cite{nnano-eichler-2011, negtuning}
\begin{equation}
 -a_s = f_\text{max}\frac{\Cg'' \ell}{4\pi^2 T_0}
\end{equation}
we obtain $\Cg'' = \text{d}^2\Cg/\text{d}x^2 = 1.0\cdot 10^{-5} \,
\text{F/m}^2$. The wire-over-plane model, with the length of the wire rescaled
as discussed above to the effective electronic length $\ell_\text{eff} =
140\un{nm}$, leads to $\Cg'' = 2.7\cdot 10^{-6}\,\text{F/m}^2$, approximately a
factor 4 smaller.

While one may typically expect symmetric behaviour around $\Vg=0\un{V}$
\cite{apl-solanki-2010}, static charges e.g.\ at the substrate surface can
explain a common offset of the extrema of both modes. This however provides no
straightforward explanation for the relative shift of the two modes in \vg, 
with the frequency maximum of the electrostatically softened mode at 
$\Vg=-2.64\un{V}$ and the minimum of the non-softened mode at $\Vg=7.48\un{V}$.

As initially shown in \cite{strongcoupling, science-lassagne-2009}, in carbon 
nanotubes Coulomb blockade effects also have a strong impact on mechanical 
resonances. A corresponding detail measurement of the two mechanical modes is
shown in Fig.~\ref{figMechanics}(d) and will be discussed below.

\section{Interaction of the subsystems}

\subsection{Microwave resonance shift due to Coulomb blockade}

\begin{figure}[t]
\begin{center}
\includegraphics{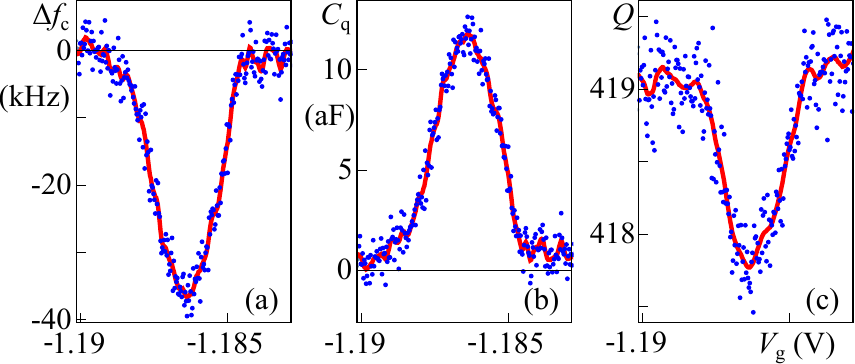}
\end{center}
\caption{
Impact of a Coulomb oscillation of conductance of the carbon nanotube quantum 
dot on the coplanar waveguide resonator: (a) resonance frequency shift $\Delta 
\fcav(\Vg)$, (b) quantum capacitance $\Cq(\Vg)$ calculated from $\Delta 
\fcav(\Vg)$, and (c) quality factor $\Qcav(\Vg)$ of the coplanar waveguide 
resonator as function of the applied static gate voltage \Vg. The red curves 
show in each case a moving average. All data points are obtained by fitting Eq. 
(\ref{eq-s21fano}) to frequency-dependent transmission measurements $\left| 
S_{21}(f) \right|^2$.\label{figFReduction}
}
\end{figure}

As already mentioned above, the effective replacement circuit capacitance 
\CRLC\ (see Fig.~\ref{figCoplanarRes}(d)) is directly related to the geometric 
capacitance \Ccav\ of the coplanar waveguide resonator, taking into account the 
spatial distribution of electric fields. For our $\lambda / 2$ resonator, we 
calculate a geometric capacitance of $\Ccav = 1750\un{fF}$, neglecting the 
small and constant coupling capacitances. In the case of the fundamental mode 
of our $\lambda/2$ resonator and its field distribution, this translates to 
$\CRLC = \Ccav / 2 = 875\un{fF}$ \cite{book-simons}. With the resonance 
frequency, we obtain a corresponding replacement circuit inductance $\LRLC = 
879\un{pH}$, which is in the following assumed constant.

The gate voltage dependent contribution of the quantum dot to the total 
replacement circuit capacitance $\Delta \CRLC(\Vg)$ can be written as quantum 
capacitance
\begin{equation}
\Delta \CRLC(\Vg) = \Cq = \frac{\partial \Qg}{\partial \Vg} = \alpha\,
\frac{\partial \Qdot}{\partial \Vg}
\end{equation}
with $\alpha$ the gate conversion factor of the quantum dot as introduced 
above. It describes the response of the gate charge \Qg\ to a gate voltage 
fluctuation \cite{prl-delbecq-2011, nature-desjardins-2017, prb-roschier-2005}. 
In Coulomb blockade, \Cq\ is effectively zero since the charge on the quantum 
dot is constant for small voltage variations. In contrast, a maximum of \Cq\ is
reached at the position of a conductance peak where the charge on the quantum 
dot varies. This becomes directly visible as a reduction of the cavity 
resonance frequency \fcav.

A corresponding measurement is shown in Fig.~\ref{figFReduction}(a). For each 
value of the gate voltage \Vg\ across a Coulomb oscillation, a trace of the 
coplanar waveguide resonator transmission $S_{21}(f)$ has been recorded. 
Fitting the transmission data with Eq.~(\ref{eq-s21fano}), we obtain the 
resonance frequency $\fcav(\Vg)$ and its change $\Delta \fcav (\Vg) = 
\fcav(\Vg) - \fcav^0$ induced by the quantum capacitance, 
Fig.~\ref{figFReduction}(a), and the resonator quality factor $\Qcav(\Vg)$, 
Fig.~\ref{figFReduction}(c). Assuming constant \LRLC, we then translate $\Delta 
\fcav \approx 36\un{kHz}$ into a change in replacement circuit capacitance 
$\Delta \CRLC(\Vg) = \Cq \approx 10\un{aF}$, see Fig.~\ref{figFReduction}(b). 
Remarkably, this effective value is larger than the bare geometric gate 
capacitance of the CNT to the gate finger $\Cg =2.6 \un{aF}$.

The quality factor $\Qcav(\Vg)$ of the microwave resonator is clearly reduced 
when the nanotube quantum dot is in single electron conduction, see 
Figure~\ref{figFReduction}(c). This effect is not covered by the circuit model, 
but can be explained as follows. As discussed below in 
Section~\ref{impact-ghz-dot} in detail, we can estimate the voltage amplitude 
of the driven cavity and with it the energy stored in the cavity. Using the 
data of Fig.~\ref{figOscPotential}(b), we obtain for the parameters of 
Fig.~\ref{figFReduction} $V\ti{ac} = 9.5\un{mV}$ and $\Ecav = 246\un{eV}$. This 
ac voltage amplitude is consistent with the width of the Coulomb oscillation in 
Fig.~\ref{figOscPotential}(a), Section~\ref{impact-ghz-dot}, at a nominal 
generator drive power of $10\un{dBm}$. 

\begin{figure}[t]
\begin{center}
\includegraphics{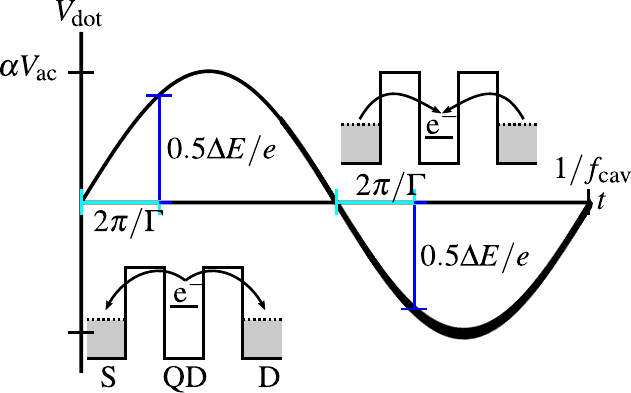}
\end{center}
\caption{Dissipation mechanism for the drop in \Qcav\ observed in 
Fig.~\ref{figFReduction}(c), see the text. Due to a finite tunnel rate 
$\Gamma$ and a corresponding delay effect, electrons are pumped from a lower to
a higher potential by the microwave signal.
\label{figPumping}
}
\end{figure}

The dip in the cavity $Q$-factor \Qcav\ indicates an additional energy loss per 
microwave period induced by single electron tunneling. The loss can be 
estimated as 
\begin{equation}
\Delta E=\Ecav \left( \frac{1}{Q\ti{c}^\text{CB}} - \frac{1}{Q\ti{c}^\text{SET}} 
\right) = 1.76\un{meV}. 
\end{equation}
Applying a model initially developed for nanoelectromechanical systems 
\cite{prb-meerwaldt-2012}, we assume that due to the finite tunnel rate
$\Gamma$ connecting the quantum dot to its leads the quantum dot occupation
only follows the potential oscillation with a delay. This way, electrons are 
pumped from lower to higher energy states, resulting in an energy loss for the 
microwave resonator. 

\begin{figure}[t]
\begin{center}
\includegraphics{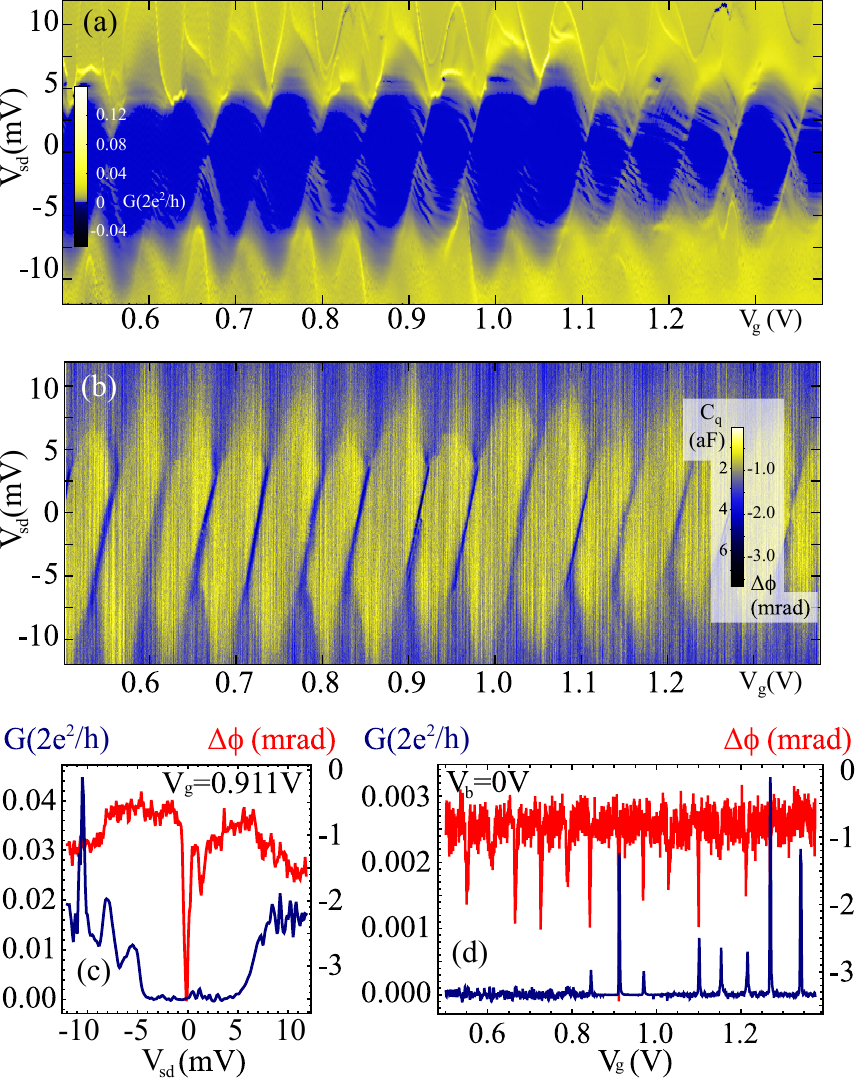}
\end{center}
\caption{(a) dc conductance and (b) simultaneously measured GHz transmission 
phase at $f = 5.73957\un{GHz}$, as function of the applied gate voltage \Vg\ 
and bias voltage \Vsd\ across the carbon nanotube quantum dot. The transmission 
was measured with a VNA input filter bandwidth of $2\un{Hz}$. (c), (d) Trace 
cuts from (a) and (b), for constant gate voltage $\Vg = 0.911\un{V}$ (c) and 
constant bias voltage $\vsd= 0 \un{V}$ (d), respectively. Dark blue / left 
axis: differential conductance; red / right axis: GHz transmission phase.
\label{figPhaseDiamonds}
}
\end{figure}

Figure~\ref{figPumping} illustrates the mechanism as well
as the expected average energy $\Delta E$ lost during one microwave period: an
electron tunnels from the quantum dot to the electrodes on average after the
time $2\pi/\Gamma$, extracting $0.5\Delta E$ from the microwave resonator. The
same amount of energy is extracted $1/\Gamma$ after the gate potential is below
the SD energy. Coulomb blockade ensures that during one microwave cycle at most 
one electron undergoes this process. The combined tunnel rate of the 
contacts $\Gamma$ thus relates to $\Delta E$ as
\begin{equation}
\frac{1}{\Gamma}=\frac{1}{\wcav} \arcsin\left(\frac{\Delta 
E}{2\alpha V_\text{ac}e}\right).
\end{equation}
For $\alpha V\ti{ac} = 2.52\un{mV}$ and the parameters given in the measurement
of Fig.~\ref{figFReduction}(c), we obtain a tunnel rate $\Gamma = 101\un{GHz}$,
in reasonable agreement with the previous estimate from Coulomb blockade 
$\Gamma = 160\un{GHz}$ for this parameter.

\subsection{Transmission phase based quantum capacitance detection}

At or close to the microwave cavity resonance \fcav, the transmission phase of 
the microwave resonator is highly sensitive to the drive frequency deviation 
$\Delta f = f - \fcav$, see Fig.~\ref{figCoplanarRes}(b). A slight shift of the 
resonance frequency, e.g., due to changes of the resonator environment, becomes 
equally visible as a transmission phase shift, with an approximate linear 
relation. This allows to efficiently probe the resonator and with it the 
quantum capacitance of the adjunct nanotube system, a technique that has already 
been applied successfully to carbon nanotubes, see \cite{prl-delbecq-2011,
nature-desjardins-2017}. Both the resonance shift (max. $40\un{kHz}$) and the 
change in resonance width from $Q$ (max. $65.5\un{kHz}$) identified in 
Fig.~\ref{figFReduction} cannot move our working point significantly relative 
to the $>10\un{MHz}$ wide cavity resonance nor significantly change the slope 
of the linear relation. With $\fcav = 5.73957 \un{GHz}$ and $\Qcav=495$ we 
obtain a phase shift of $\phi/\Delta f= 0.144 \un{mrad/kHz}$.

A corresponding measurement is shown in Fig.~\ref{figPhaseDiamonds}. 
Fig.~\ref{figPhaseDiamonds}(a) and Fig.~\ref{figPhaseDiamonds}(b) plot the 
simultaneously measured dc conductance and microwave transmission phase, as 
function of applied gate voltage \Vg\ and bias voltage \Vsd, over a range of 
several Coulomb oscillations. Fig.~\ref{figPhaseDiamonds}(c) and 
Fig.~\ref{figPhaseDiamonds}(d) are trace cuts from the measurement, for (c) 
constant gate voltage $\Vg=0.911\un{V}$ and (d) constant bias voltage $\Vsd=0$.

The Coulomb oscillations and with them the oscillatory behaviour of the 
quantum capacitance in \Vg\ are immediately visible. In 
Fig.~\ref{figPhaseDiamonds} we use an input filter bandwidth of the VNA of 
$\Delta f_\text{IF}=2\un{Hz}$, corresponding to an integration time per point 
on the order of $0.5\un{s}$. From the phase noise $\Delta \varphi \sim 
0.2\un{mrad}$ of the trace cut of Fig.~\ref{figPhaseDiamonds}(c) we can 
estimate a measurement sensitivity of $2\un{aF}$ or better, see also 
\cite{nature-desjardins-2017}. While this does not reach the resolution of 
Fig.~\ref{figFReduction}(b), it is recorded significantly faster.

In Fig.~\ref{figPhaseDiamonds}(b), the phase shift highlights a preferred edge 
of the single electron tunneling regions as the parameter region where the 
time-averaged charge of the quantum dot changes by one electron. This indicates 
that the tunneling rates from the quantum dot to source and drain differ 
significantly; in single electron tunneling the time-averaged charge is close 
to one of the neighbouring Coulomb blockade regions. Charging of the quantum 
dot predominantly happens when the quantum dot potential crosses the Fermi edge 
of the contact with the higher tunneling rate. Similar observations have 
already been made on quantum dots with asymmetrically coupled reservoirs 
\cite{prb-schleser-2005, nature-desjardins-2017}; for an in detail analysis 
of charging and tunnel rates see \cite{prb-schleser-2005}, where a quantum 
point contact charge detector is used to obtain an equivalent signal.

\subsection{Impact of GHz signals on the quantum dot}\label{impact-ghz-dot}

\begin{figure}[t]
\begin{center}
\includegraphics{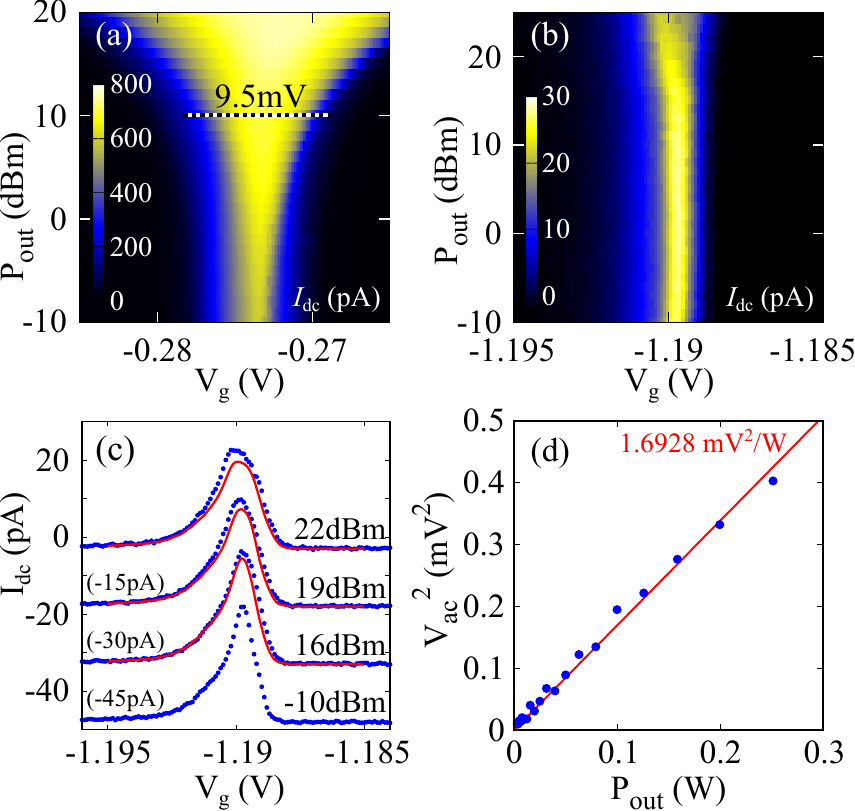}
\end{center}
\caption{(a,b) Broadening of a Coulomb oscillation while an (a) resonant and 
(b) off resonant microwave drive at differing output power is applied to the 
microwave resonator: dc current $I_\text{dc}$ as function of gate voltage \Vg\ 
and rf generator output power $P_\text{out}$. (a) $\Vsd=2\un{mV}$, $\fdrive =
5.73957\un{GHz}$; (b) $\Vsd=0.15\un{mV}$, $\fdrive = 5.23989\un{GHz}$; data 
already shown in \cite{optomechanics}, Fig. S10(a).
(c) Data points: trace cuts $I_\text{dc}(\Vg)$ from (b) at $P_\text{out} = 
22\un{dBm}, 19\un{dBm}, 16\un{dBm}, -10\un{dBm}$. Solid lines: Fits of the
numerically broadened low-power trace at $-10\un{dBm}$, resulting in a 
modulation ac voltage $V\ti{ac}$ (see the text for the model). The different 
drive powers are offset for clarity. (d) Plot of the square of the 
modulation voltage $V\ti{ac}^2$ extracted from the data of (b), as function of 
the applied generator output power $P_\text{out}$, and linear fit, resulting in 
a proportionality factor $1.6928\un{mV$^2$/W}$; data already shown in 
\cite{optomechanics}, Fig. S10(b).
\label{figOscPotential}
}
\end{figure}

The impact of a microwave signal in the resonator on dc (or, more precisely, 
time-averaged / rectified) transport through the quantum dot, i.e., the reverse 
effect compared to above discussion, is shown in Fig.~\ref{figOscPotential}. 
Since the electronic tunnel rates and thus the Coulomb oscillation width 
$\Gamma\ti{est}
\simeq 163\,$GHz \cite{optomechanics} clearly exceed the drive frequency
$\fdrive \simeq 5\un{GHz}$, we can treat the microwave signal as a classical
oscillating gate voltage. In a first approximation, this signal, too fast for 
our low-frequency circuit to follow, thus effectively widens the observed 
Coulomb oscillations. 

This is demonstrated in Fig.~\ref{figOscPotential}(a) for a resonant and in 
Fig.~\ref{figOscPotential}(b) for an off-resonant cavity drive. In 
Fig.~\ref{figOscPotential}(a), increasing $P\ti{out}$ additionally leads to a
peak current increase, indicating that at the resulting large photon numbers in 
the cavity the approximation of broadening of the peak only breaks down. For 
the off-resonant case in Fig.~\ref{figOscPotential}(b), the peak current 
decreases with applied power, and we can model the impact of the ac signal 
numerically by averaging a (near) zero-drive gate trace over a sinusoidal gate 
voltage of given amplitude. 

In detail, we extract a trace $I_0(\Vg)$ at small or zero drive amplitude 
(here, at $P_\text{out}=-10\un{dBm}$) and then numerically find the ac gate 
voltage amplitude $\Vg^\text{ac}$ such that the average
\begin{equation}\label{eq:broadenedcb}
I_\text{driven}(\vg) = \frac{1}{2\pi} \int\limits_0^{2\pi}
I_0\left(\vg + \vgac\sin(\varphi)\right)\, \text{d}\varphi
\end{equation}
best fits to a measured trace $I(\Vg)$ at finite drive amplitude. Example 
results are shown in Fig.~\ref{figOscPotential}(c), for data measured at 
$P_\text{out} = 22\un{dBm}, 19\un{dBm}, 16\un{dBm}$ and the resulting best ac 
voltages for the broadening $\vgac=0.53\un{mV}, 0.37\un{mV}, 0.25\un{mV}$. 
The lowermost set of points shows the reference trace at $P_\text{out} = 
-10\un{dBm}$. Fig.~\ref{figOscPotential}(d) plots the square of the ac 
voltage as function of generator power $P_\text{out}$, demonstrating that these 
values are proportional as expected.

We approximate that due to the proximity of the nanotube transfer regions to 
the coupling capacitors of the coplanar waveguide resonator the ac gate voltage 
amplitude is equal to the voltage amplitude at the resonator antinode (i.e., at 
the coupling capacitor). This allows an estimate of the photon number in the 
resonator as function of applied drive power. Using the linear fit of
Fig.~\ref{figOscPotential}(d) (red sideband drive) and the replacement circuit capacitance $\CRLC = 
875\un{fF}$ as discussed above, a generator drive power of $P_\text{out} = 
0.1\un{W} = 20\un{dBm}$ translates to a voltage amplitude of $\vgac \simeq
0.41\un{mV}$ and to $\ncav \simeq 21300$ resonator photons. A more detailed
discussion of error sources for the estimation can be found in
Appendix~\ref{app-N}.

Note that the measurements of Fig.~\ref{figOscPotential}(b-d) have been
performed with an off-resonant drive of the cavity. In the case of a resonant
drive, the increased photon occupation (larger by a factor of $|S_{21}(f_0)/S_{21}(f)|^2$) leads to a 
correspondingly stronger ac signal and thereby broadening, see 
Fig.~\ref{figOscPotential}(a). Calculating the expected peak width for 
$P\ti{out} = 10\un{dBm}$ at resonance gives $9.5\un{mV}$, which is in good 
agreement with the observation in Fig.~\ref{figOscPotential}(a), see the scale 
bar in the figure. Additionally, the maximum current increases at large power, 
which is in clear disagreement with our simple broadening model. For tunneling 
through a single, discrete level in the quantum dot, heating of the electron 
gas in the contacts leads to a decrease of the current 
\cite{prb-beenakker-1991}. This leaves as explanation for the increase either 
accessing excited states as additional transport channels via the broadened 
Fermi distribution of the contacts or more complex processes such as 
(multi-)photon-assisted tunneling.

\subsection{Interaction of vibration and Coulomb blockade}

Figure~\ref{figMechanics}(d) shows a detail measurement of the driven
mechanical resonances, corresponding to a zoom of the region marked in
Fig.~\ref{figMechanics}(a) with a gray bar. Here, the impact of the Coulomb
oscillations on the two mechanical modes becomes clearly visible 
\cite{strongcoupling}. Comparison with Fig.~\ref{figMechanics}(a) lets us 
conclude that the lower mode with $499.4\un{MHz} \le f \le 500\un{MHz}$ is 
predominantly the globally softening (perpendicular to the device surface) mode 
and the upper mode with $500.5\un{MHz}\le f \le 500.3\un{MHz}$ is predominantly 
the globally hardening (parallel to the device surface) mode. The measurement 
is still near the mode anticrossing at $\Vg\simeq 7.5\un{V}$, however, such 
that a finite mode mixing cannot be excluded.

The dashed lines in Fig.~\ref{figMechanics}(c) correspond to fits to the 
extracted resonance positions. We simplify the Coulomb oscillations as a 
sequence of 6 equidistant Lorentzians in conductance, with the corresponding 
increase of the time-averaged number of electrons in the quantum dot $\left< N 
\right>\!(\Vg)$ from the resonant tunneling picture. Since the oscillation at 
$\Vg\simeq 8.88\un{V}$ does not fit in to this equidistant peak scheme (which 
is not particularly surprising for a carbon nanotube quantum dot with 
significant quantum mechanical contributions to the addition energy), we ignore
the data points around this one Coulomb oscillation. The overall fit function
used for each of the two mechanical modes ($i=1,2$) is
\begin{equation}
f_i(\Vg)= a_{1,i} + a_{2,i}\Vg + a_{3,i} \left< N \right>(\Vg) + a_{4,i} 
\frac{\text{d}\left< N \right>}{\text{d}\Vg}
\end{equation}
where in particular $a_{4,i}$ captures the {\em local} electrostatic softening 
through the Coulomb blockade oscillation, expressed in \cite{strongcoupling} 
for the spring constant as
\begin{equation}
a_{4,i} \propto \Delta k \propto \left( \frac{\text{d}\Cg}{\text{d} x} 
\right)^2
\end{equation}
(Eqn.~(S5) there, with other control parameters such as \Vg\ constant).

For a vibration mode whose motion does not change the gate capacitance, no 
local capacitive softening is expected, and indeed the higher frequency mode 
in Fig.~\ref{figMechanics}(c) shows a much smaller coupling to the Coulomb 
oscillations. From the fits, we obtain a ratio $a_{4,2}/a_{4,1} = 0.29$, 
leading to a ratio of the capacitance sensitivities $(\text{d}C_g/\text{d}x_2) 
/ (\text{d}C_g/\text{d}x_1) = \sqrt{0.29} \simeq 0.54$, which would be 
fulfilled for two relatively perpendicular vibration modes both rotated by 
$\arctan(0.54) \simeq 30^\circ$ to the device surface normal. For future 
measurements it would be interesting to trace a mode anticrossing as in 
Fig.~\ref{figMechanics}(a) in more detail and extract the evolution of the 
couplings and their ratio as function of gate voltage in this 
nano-electromechanical model system.

\section{Quantum capacitance enhanced optomechanics}

\subsection{Introduction}

Dispersive coupling is both experimentally and theoretically the most widely 
researched mechanism for obtaining an optomechanical system 
\cite{rmp-aspelmeyer-2014}. Here, mechanical displacement causes a change in
resonance frequency of an electromagnetic / optical resonator. In a microwave 
optomechanical system, this typically happens via a modification of the 
capacitance of a $LC$-circuit; one of the capacitor electrodes is the 
mechanically active element.

\begin{figure}[t]
\begin{center}
\includegraphics{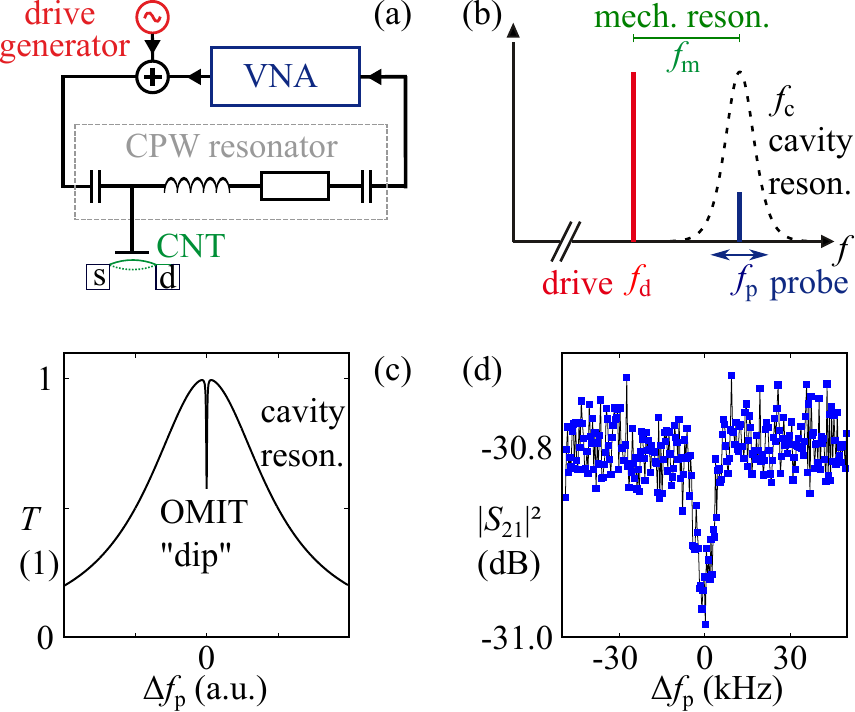}
\end{center}
\caption{
(a) Circuit schematic of our optomechanically induced transparency (``OMIT'') 
experiment (simplified from \cite{optomechanics}). (b) Signal frequency 
schematic. (c) Schematic plot of the cavity transmission $T= \left| S_{21} 
\right|^2$ (using Eq.~\ref{intracav}) as function of $\Delta \fprobe = \fprobe 
- 
\fcav$ for an OMIT experiment as in (a), with $\fdrive=\fcav-\fmech$. (d) 
Measurement of the OMIT ``dip'' in transmission, for $\vg = -1.18751\un{V}$  
(cf. (c); trace from \cite{optomechanics} Fig. 2(f)).
\label{figOMIT}
}
\end{figure}

While optomechanical effects in a combined nanotube--microwave resonator system
as shown in Fig.~\ref{figOMIT}(a) can in principle occur due to geometrical 
capacitance changes alone, a quick estimate already shows that the resulting 
coupling parameters are tiny \cite{optomechanics}. Essentially, this is a 
manifestation of a mismatch of scales. Working frequencies for coplanar 
waveguide resonators are typically in the range $4-8\un{GHz}$. With this one 
obtains wavelengths and resonator sizes on the order of $\sim 1\un{cm}$. A 
carbon nanotube segment with ballistic conduction, where electrons can be 
confined to single, well-separated and unperturbed quantum levels, has 
typically a length of $\lesssim 1 \, \mu \text{m}$. Mechanical deflections of 
such a segment as vibrational resonator are in the range of (at strong driving) 
$\sim 1\un{nm}$ \cite{magdamping} and (zero point motion) $\sim 1\un{pm}$ 
\cite{optomechanics}. Obviously, a deflection $\lesssim 1\un{nm}$ will barely 
affect the geometric properties of a resonator of size $1\un{cm}$.

In the following, the nonlinear charging characteristic of the quantum dot
embedded in the nanotube is used for amplification of the coupling. As 
demonstrated above, it leads to a gate voltage dependent, locally strong 
quantum capacitance contribution, which can dominate geometric effects. The 
mechanical oscillation is slow compared to both tunnel rates and microwave 
resonance frequency, allowing us to treat this quantum capacitance as a 
replacement parameter which again depends on the deflection. At proper choice 
of the gate voltage working point, optomechanical experiments become possible 
\cite{optomechanics}.

\begin{table}
\begin{center}
\begin{tabular}{|p{4cm} |c|c|c|}
	\hline 
	\textbf{Input signals} & & & \\
	Drive generator power & & $P_\text{d}$ & $25\un{dBm}$ \\
	Probe (VNA) power & & $P_\text{p}$ & $-20\un{dBm}$ \\
	estimated cable damping & & $\left|S_c\right|^2$ & $8\un{dB}$ \\
	\hline
	\textbf{Subsystems} & & & \\
	Cavity resonance frequency &  &  \fcav & $5.74005\un{GHz}$ \\
	Cavity line width &  & $\kcav$ & $2\pi\cdot 11.55\,\text{MHz}$  \\
	Cavity quality factor &  & $\Qcav$ & $497$  \\
	Mech. resonance frequency & & $\fmech$ & $502.536\,\text{MHz}$ \\
	Mech. line width & & $\kmech$ & $2\pi\cdot 5.884\un{kHz} $ \\
	Mech. quality factor & & $\Qmech$ & $85407$ \\
	\hline
	\textbf{Coupling parameters}\newline
	($\ncav=67500$) & & & \\
	Side band resolution & $2\pi\fmech/\kcav$ & & 43.5 \\
	Single photon coupling & & $g_{\text{0}}$ & $2\pi\cdot 94.2 \,
\text{Hz}$
		\\
	Optomechanical coupling & $ g_{\text{0}}\sqrt{\ncav}$ & $g$ &
		$2\pi\cdot 24.47 \un{kHz}$ \\
	Cavity pull-in parameter & $g_0/\xzpf$ & $G$ & $2\pi\cdot 50.5 
		\un{Hz/pm}$ \\
	Dispersive coupling & ${g_0}/{\kcav}$ & $\tilde{A}$ &	$8.16 \cdot
            10^{-6}$ \\
	Max. sideband cooling rate & $4\ncav
		g_{\text{0}}^2 /{\kcav}$ & $\Gamma_{\text{opt}}$ & $2\pi\cdot
		207.4 \un{Hz}$ \\
	Cooperativity & ${\Gamma_{\text{opt}}}/{\kmech}$ & $C$ & $0.035$ \\
	Cooling power & $\Gamma_{\text{opt}} h \fmech$ & $\dot{Q}$ &
            $4.34\cdot 10^{-22}\,\text{W}$  \\
	\hline
\end{tabular}
\end{center}
\caption{
Overview of the setup and optomechanical device parameters, corresponding to the
measurements of Fig.~\ref{figOMIT}, Fig.~\ref{figCoupling}, and
Fig.~\ref{figDamping}. Except for the cavity parameters evaluated separately,
the values originate from the OMIT transmission curve fits and correspond to
the data points in Fig.~\ref{figCoupling} and Fig.~\ref{figDamping} for
$\Vg=-1.18754\un{V}$.
\label{parameters}
}
\end{table}

\subsection{Optomechanically induced (in)transparency OMIT}

Our experiment for detecting and determining the optomechanical coupling is a
so-called optomechanically induced transparency (``OMIT'') measurement,
introduced first by Weis {\it et al.} \cite{science-weis-2010} and based on
earlier work on electromagnetically induced transparency of resonator media
\cite{prl-boller-1991}. Fig.~\ref{figOMIT}(a) displays the simplified high
frequency circuit, Fig.~\ref{figOMIT}(b) schematically sketches the involved
frequencies. A strong drive signal \fdrive\ is applied at a constant frequency
red-detuned from the cavity resonance by the mechanical resonance frequency, 
$\fdrive = \fcav - \fmech$. Additionally a probe signal \fprobe\ is swept 
across the cavity resonance and its transmission measured.

Extending Eq.~(\ref{eq-s21fano}), the transmission of the probe signal,
proportional to the intracavity photon number, now follows the broad peak of
the electromagnetic resonance everywhere except near $\fprobe = \fdrive +
\fmech$, where a dip in transmission emerges, see also Fig.~\ref{figOMIT}(c):
\begin{align}\label{intracav}
S_{21}(\fprobe)=\frac{2A}{4\pi i(\fcav-\fprobe)+
\Gacav+
\frac{4 g^2}{4\pi i(\fmech+
\fdrive-\fprobe)+\Gamech}}+re^{i\theta}
\end{align}
Here, $g=\sqrt{\ncav}\,g_0$ is the optomechanical coupling for the total
number of cavity photons $\ncav$, which is at fixed drive frequency proportional
to the drive power. Microscopically Eq.~(\ref{intracav}) can be motivated such
that mechanical phonons of frequency \fmech\ pair up with photons of \fdrive,
upconverting to the cavity frequency $\fcav$, and then interfere destructively
with the probe signal at $\fprobe$ \cite{science-weis-2010}, suppressing the 
population of the cavity.

The width of the optomechanically induced transmission dip in
Eq.~(\ref{intracav}) is given by the effective mechanical damping rate
\Gameff\ \cite{rmp-aspelmeyer-2014},
\begin{equation}\label{effectivedamping}
\Gameff = \Gamech +  \Gamma_{\text{opt}},
\end{equation}
where the latter term in the sum, see also Table~\ref{parameters},
\begin{equation}
\Gamma_{\text{opt}}=\frac{4g^2}{\Gacav},
\end{equation}
is exactly the dispersive optomechanical damping (or cooling) rate induced by 
the red sideband detuned drive signal in absence of the weak probe signal.

\subsection{Quantum capacitance amplified coupling}

\begin{figure}[t]
\begin{center}
\includegraphics{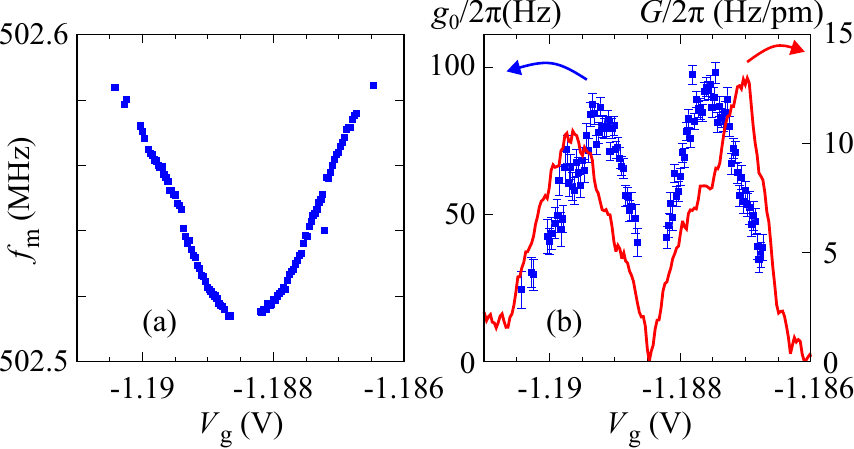}
\end{center}
\caption{OMIT-derived parameters across a Coulomb blockade oscillation, as a 
function of gate voltage \Vg: (a) mechanical resonance frequency \fmech, (b) 
blue points, optomechanical single photon coupling $g_0(\Vg)$ (left axis, 
data already shown in \cite{optomechanics}, Fig. 2(g)); red 
line, cavity pull-in parameter $G(\Vg)$ derived from the quantum capacitance 
$C_\text{q}(\Vg)$ of Fig.~\ref{figFReduction}(b) for comparison (right axis).
\label{figCoupling}
}
\end{figure}
As visible from Eq.~(\ref{intracav}), the multi-photon optomechanical coupling
$g(\Vg)$ as well as several other parameters can be extracted directly from
OMIT measurements as curve fit parameters. This becomes particularly 
interesting when stepping the dc gate voltage \Vg\ across a Coulomb oscillation 
and plotting the parameters as function of gate voltage. Already well-known 
behaviour can be seen in Fig.~\ref{figCoupling}(a), with the gate voltage 
dependence of the mechanical resonance frequency $\fmech(\Vg)$, now extracted 
from fits of Eq.~(\ref{intracav}) to OMIT data. The decrease of $\fmech(\Vg)$ 
corresponds to the local electrostatic softening (or the ``Coulomb oscillations 
of mechanical resonance frequency'') as also already demonstrated in
Fig.~\ref{figMechanics}(d) and in earlier publications \cite{strongcoupling,
highqset, kondocharge}.

Using the cavity photon number \ncav\ as determined previously, see 
Fig.~\ref{figOscPotential}, the single photon optomechanical coupling $g_0(\vg) 
= g(\vg) / \sqrt{\ncav}$ can be calculated; it is plotted in
Fig.~\ref{figCoupling}(b) (blue points, left axis). The data clearly shows 
maxima on the flanks of the Coulomb oscillation, while the coupling both 
vanishes at its center and in Coulomb blockade. This behaviour has already been 
discussed in detail in \cite{optomechanics}. There it was shown that $g_0(\vg)$ 
is connected to the time-averaged number of electrons on the quantum dot 
$\avn\!(\vg)$, increasing by one over a Coulomb oscillation, via 
\begin{equation}\label{enhancedg0}
 g_0(\vg) = \left| \frac{\pi e \alpha\fcav \Vg}{\CRLC \Cg}
        \,\frac{\partial \Cg}{\partial x}\,
        \frac{\partial^2 \avn}{\partial \Vg^2} \right| \,\xzpf
\end{equation}
Assuming a lifetime-broadened level in the quantum dot and thus a Lorentzian
shape of ${\partial \avn}/{\partial \Vg}$, Eq.~(\ref{enhancedg0}) allows to
approximate the functional dependence of the $g_0(\vg)$ data in
Fig.~\ref{figCoupling}(b) (points) very well. In quantitative terms, theory and 
experiment differ by approximately a factor 5, still an excellent agreement 
given the amount of approximations that enter the calculation 
\cite{optomechanics}.

An alternative way to characterize the optomechanical coupling is directly via
the cavity pull-in parameter $G(\Vg)$, i.e., the cavity resonance frequency 
\fcav\
shift per mechanical displacement,
\begin{equation}
G  = \frac{g_0}{\xzpf} = 2\pi \frac{\partial\fcav}{\partial x}.
\end{equation}
We can extract this parameter from the gate voltage dependence of the resonance
frequency $\fcav(\vg)$, Fig.~\ref{figFReduction}(a), which directly provides us
the quantum capacitance $\Cq(\vg)$, Fig.~\ref{figFReduction}(b). The details
are given in Appendix~\ref{app-G} and lead to
\begin{equation}\label{eq:Gfrommeas}
 G = \frac{\wcav}{2\CRLC} \, \frac{\partial \Cq}{\partial \Vg} \,
 \frac{\Vg}{\Cg} \, \frac{\partial \Cg}{\partial x}
\end{equation}
where again \Cg\ is the geometric capacitance between resonator (gate) and
nanotube (quantum dot), which can be extracted rather precisely from Coulomb
blockade measurements, and \Cq\ is the quantum capacitance as plotted in 
Fig.~\ref{figFReduction}(b).

The result for $G(\Vg)$ is plotted in Fig.~\ref{figCoupling}(b) as a solid red
line. An offset in gate voltage has been corrected here; it most likely was 
caused by charging effects over the course of the lengthy measurement 
cool-down. The functional dependence agrees well with the OMIT result. The
Coulomb oscillation structure appears to be slightly wider in gate voltage
for $G(\Vg)$. Comparing the applied powers of the resonant probe signal in the 
OMIT case $P_\text{p} = -20\un{dBm}$ and the repeated cavity resonance sweeps of
Fig.~\ref{figFReduction}, $P= +10\un{dBm}$, this effect is however well within 
the possible broadening of the Coulomb oscillation by the GHz signal.

Since the two parameters $g_0$ and $G$ are proportional and their
proportionality factor $G \xzpf=g_0$ is exactly the zero point fluctuation 
scale of the mechanical system, this provides us a way to estimate \xzpf. 
Bringing the two curves in Fig.~\ref{figCoupling}(b) to best agreement leads to 
$\xzpf = 7.4\un{pm}$. Using the harmonic oscillator expression $\xzpf =
\sqrt{\hbar/2m\wmech}$ with the effective mass as given in Table~\ref{par-cnt}
and used otherwise in the calculations, we obtain $\xzpf = 1.9\un{pm}$; again
the values agree better than one order of magnitude.

Regarding error sources, for the cavity pull-in parameter $G(\Vg)$,
Eq.~(\ref{eq:Gfrommeas}), all values can be directly read out from the
measurement, with the exception of $\partial\Cg / \partial x$. The latter is
calculated by scaling the wire-over-plane model for a gated carbon nanotube
\cite{optomechanics, apl-wunnicke-2006} down to the effective electronic 
length $\ell_\text{eff}=140\un{nm}$ of the nanotube quantum dot: we calculate 
the theoretical capacitance between suspended nanotube and gate from the device 
geometry, compare it with the Coulomb blockade derived gate capacitance \Cg\ to 
obtain an effective electronic length, and scale the theoretical derivative 
$\partial\Cg/\partial x$ accordingly. 

\begin{figure}[t]
\begin{center}
\includegraphics{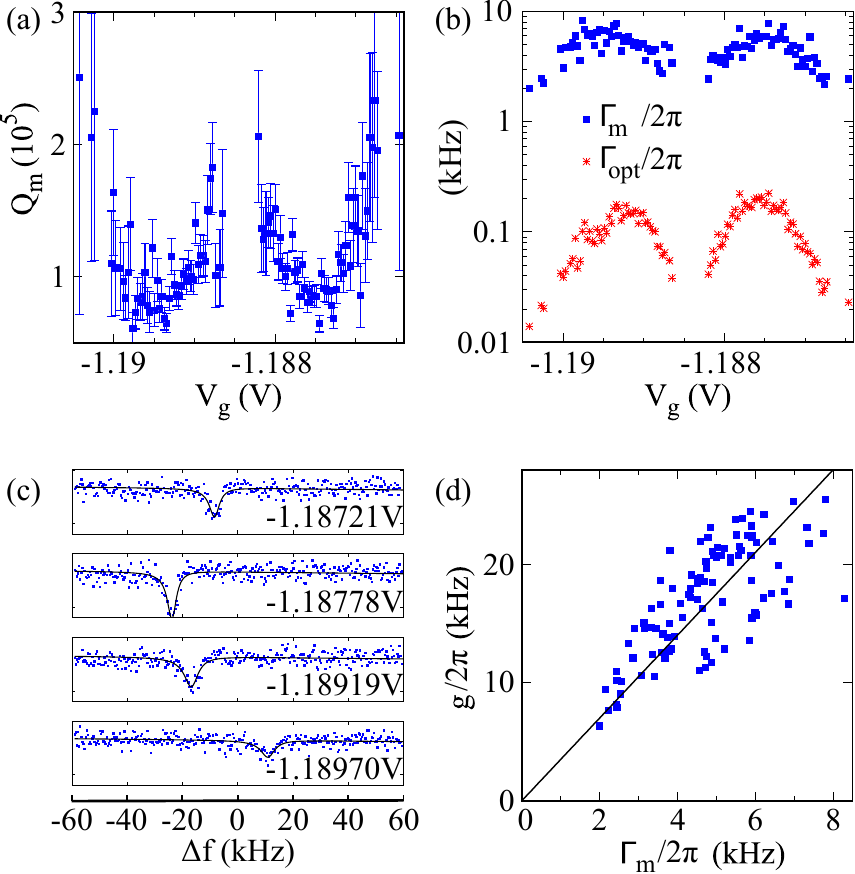}
\end{center}
\caption{(a) Mechanical quality factor $\Qmech = 2\pi \fmech/\Gamech$ observed 
during the OMIT experiment, as function of gate voltage \vg.
(b) Mechanical damping rate \Gamech\ (blue squares) and optomechanical damping 
rate $\Gamma_\text{opt}=4g^2/\Gacav$ (red stars) from the OMIT measurement.
Both $g$ and $\Gamech$ habe been obtained via fitting Eqn.~\ref{intracav}.
(c) Example raw data traces of the OMIT measurement, showing the OMIT ``dip''
in transmission $\left| S_{21}(f) \right|^2$ at different gate voltages.
(d) Optomechanical coupling $g$ plotted as function of the corresponding
observed damping $\Gamech$. The solid line is a linear fit $g=a\Gamech$ to all 
data points, resulting in  $a=3.5$.
\label{fig:damping}\label{figDamping}
}
\end{figure}

Conversely, the main error source for the
single photon optomechanical coupling $g_0(\Vg)$ from \cite{optomechanics} is 
the cavity photon number \ncav, derived with the assumption that the gate 
electrode shows the same GHz voltage amplitude as the end of the coplanar 
waveguide resonator at its coupling capacitance (and voltage antinode). This 
error could be reduced by, e.g., more detailed finite element modeling of the 
device. However, since the exact position and orientation of the 
deposited carbon nanotube is unknown in the experiment, it is unclear whether 
the additional effort would be of much help.

\subsection{Damping of the motion in OMIT}

Figure \ref{fig:damping}(a) plots the mechanical quality factor \Qmech\
extracted from OMIT via Eq.~(\ref{intracav}). \Qmech\ displays two distinct 
minima on the Coulomb oscillation flanks, and on the whole a behaviour inverse 
to the optomechanical coupling $g = \sqrt{\ncav} g_0$. This is clearly 
different from damping induced by electronic tunneling alone, see, e.g., 
Meerwaldt {\it et al.} \cite{prb-meerwaldt-2012}, where for $\vsd \simeq 0$ 
only a single minimum of the quality factor is observed.

The total, effective damping rate of the mechanical system taking into account 
optomechanical coupling is given by Eq.~(\ref{effectivedamping}), combining 
mechanical behaviour and the damping via upconversion of the red-detuned drive 
signal. For comparison of scales, the extracted damping rate \Gamech, inversely 
proportional to \Qmech, varies in the range $2\un{kHz}\le \Gamech/(2\pi) \le 
9\un{kHz}$. At the same time, we find $8\un{Hz} \le \Gamma_\text{opt}/(2\pi) 
\le 215\un{Hz}$. This indicates that even at our enhanced optomechanical 
coupling the damping via upconversion from Eq.~(\ref{effectivedamping}) is 
small in the experiment. Figure \ref{fig:damping}(b) plots both values as
function of \vg, confiming this conclusion with a nearly two orders of
magnitude smaller optomechanical damping for any gate voltage.

An apparent variation in \Qmech\ and the width of the OMIT dip can be caused by
mechanically nonlinear behaviour. Equation~(\ref{intracav}) assumes a harmonic
oscillator; if strong driving leads to a distortion of the mechanical resonance
shape towards a Duffing curve, the fit will return artificially smaller \Qmech\ 
values. Fig.~\ref{fig:damping}(c) plots several raw data curves of the
frequency-dependent power transmission as examples. While occasionally
asymmetric curve shapes can be observed in the raw data with its scatter, no 
systematic gate voltage dependence of this nonlinear behaviour emerges. Thus, 
no conclusion about the impact of nonlinearity on the fit results can be made.

In Fig.~\ref{fig:damping}(d), the data behind Fig.~\ref{fig:damping}(a) are
plotted showing $g(\Gamech)$, i.e., the optomechanical coupling $g$ as
function of the mechanical damping \Gamech\ (extracted from the fits); the plot 
indicates a possible linear relation between the two parameters. The cause of 
this linear relation is so far unknown; it may be due to a more complex 
interaction of Coulomb blockade, mechanics, and the microwave fields. A 
large body of theoretical literature and also experiments on the interaction of 
coherent superconducting qubit systems and optomechanical systems exists, see 
also below. However, the detailed properties differ from our single electron 
tunneling case, such that e.g.\ results from \cite{ncomms-pirkkalainen-2015} on 
the damping cannot be directly transferred.

\section{Conclusions and outlook}

A piece-by-piece characterization of a novel optomechanical device has been
presented, combining a suspended carbon nanotube as quantum dot and mechanical 
resonator with a superconducting coplanar microwave resonator 
\cite{optomechanics}. The properties of the separate three subsystems have been 
discussed in detail, as also their pairwise interactions. This includes the 
dispersion of the observed mechanical modes and their interaction with single 
electron tunneling, as well as the direct impact of the nanotube quantum dot on 
the coplanar resonator transmission phase and damping. Subsequently, the 
combined device has been introduced as a quantum capacitance enhanced 
optomechanical system. Its properties as already shown in \cite{optomechanics} 
are presented and the discussion is extended significantly. 

An alternative evaluation based on measuring the full cavity transmission 
curve allows to estimate the zero point motion scale \xzpf\ of the nanotube, 
with the result well within expected range. Further, the gate voltage 
dependence of the damping of the mechanical system during an OMIT experiment
is extracted. We find that the observed functional dependence cannot be 
explained by Coulomb blockade, nanoelectromechanical interaction, or 
optomechanics alone, indicating a more complex mechanism. Different evaluation 
paths of the measurements on device and subsystems lead to near-equivalent 
results, indicating a high degree of consistency of our total data set. 

Starting from the initial experiment combining transmon qubit and a cavity
with a nanomechanical resonator \cite{nature-pirkkalainen-2013}, much
theoretical \cite{pra-pflanzer-2013, prl-heikkila-2014, njp-rimberg-2014, 
prl-abdi-2015, prl-gramich-2013, pra-khan-2015} and experimental work 
\cite{ncomms-pirkkalainen-2015, cphys-schmidt-2020} has been invested 
worldwide in similar {\em superconducting} systems. This includes generic qubit 
treatment, but also specifically the importance of the Josephson inductance 
\cite{prl-heikkila-2014, njp-rimberg-2014} and the Josephson capacitance 
\cite{prb-manninen-2022}. Damping mechanisms are experimentally analyzed in 
\cite{ncomms-pirkkalainen-2015}. However, given the sequential electronic 
tunneling in our {\em normal-conducting} carbon nanotube quantum dot, it is 
not a priori clear inhowfar these discussions apply to our work. They are 
certainly closer to the situation of a double quantum dot as charge qubit, see, 
e.g., also \cite{prx-pistolesi-2021}, and would for sure be relevant for a 
carbon nanotube as weak link modulating an optomechanical system via the 
Josephson inductance \cite{prl-heikkila-2014, njp-rimberg-2014, 
nres-kaikkonen-2020}.

Given that the nanotube motion affects both the cavity resonance frequency and
the cavity linewidth, see Fig.~\ref{figFReduction}, a remaining open question is
whether additional dissipative optomechanical coupling \cite{prl-elste-2009,
prl-elste-erratum-2009, prl-li-2009, njp-weiss-2013} plays a role here. From a
theoretical viewpoint, the central advantage of dissipative optomechanical
coupling is that it does not require the ``good cavity limit'' $2\pi \fmech \gg
\Gacav$ for eventual ground state cooling of the mechanical system
\cite{prl-elste-2009}; in the present experiment with $2\pi \fmech / \Gacav
\simeq 44$ however this limitation of dispersive coupling is not relevant. In
addition, the prototypical dissipative system of \cite{prl-elste-2009,
prl-elste-erratum-2009} assumes an overdamped cavity where coupling to the
drive port limits $\Qcav$, an assumption far from the device parameters of our
strongly underdamped resonator with maximum transmission below $-50\un{dB}$.
This is a topic which can be addressed in future better suited devices.

Regarding further future research, the obvious path is to improve the coupling
and subsystem parameters; given the already surprising results of
Table~\ref{parameters} (updated with respect to \cite{optomechanics} to reflect 
the more precise evaluation of the mechanical resonance), reaching strong 
optomechanical coupling is likely within realistic technological reach.
Sharper Coulomb oscillations via a lower electronic tunnel rate $\Gamma$ (and 
possibly lower electronic temperature and better filtering of voltage 
fluctuations) can compress the charge increase into a smaller potential range. 
This will increase the quantum capacitance-mediated coupling $g_0$, as long as 
the separation of time scales via $\Gamma \gg 2\pi\fcav$ holds. Ongoing work
targets the circuit geometry \cite{stepwisefab}, adding on-chip filters to 
avoid GHz leakage through the dc contacts and thus achieve larger resonator 
quality factors $\Qcav$. Further options which may be considered in the future 
include an entirely changed circuit layout and the use of high kinetic 
inductance materials to maximize the impact of a changing capacitance at 
constant resonance frequency $\fcav$.

Physically, the time-dependent evolution of the coupled system is certainly a 
worthwhile object of investigation, as are coherence effects comparing single- 
and multi-quantum dot systems, and implications of more complex optomechanical 
coupling mechanisms \cite{pra-khan-2015, pra-xiong-2016, pra-hu-2015}.

\section*{Acknowledgments}

The authors acknowledge funding by the Deutsche Forschungsgemeinschaft via
grants Hu 1808/1 (project id 163841188), Hu 1808/4 (project id 438638106), Hu 
1808/5 (project id 438640202), SFB 631 (project id 5485864), SFB 689 (project 
id 14086190), SFB 1277 (project id 314695032), and GRK 1570 (project id 
89249669). A.~K.~H. acknowledges support from the Visiting Professor program of 
the Aalto University School of Science. We would like to thank O.~Vavra, 
F.~Stadler, and F.~Özyigit for experimental help, P.~Hakonen for insightful 
discussions, and Ch.~Strunk and D.~Weiss for the use of experimental facilities. 
The data has been recorded using Lab::Measure\-ment \cite{labmeasurement}.

\section*{Author contributions}

A.~K.~H. and S.~B. conceived and designed the experiment. P.~S. and R.~G.
developed and performed nanotube growth and transfer; N.~H. and S.~B.
developed and fabricated the coplanar waveguide device. The low temperature 
measurements were performed jointly by all authors. Data evaluation was done 
jointly by S.~B., N.~H., and A.~K.~H. The manuscript was written by N.~H. and 
A.~K.~H. with help from A.~N.~L.; the project was supervised by A.~K.~H.

\appendix

\section{Photon number}\label{app-N}

The estimation of the resonator photon number in Section~\ref{impact-ghz-dot}
has two limitations. On the one hand, it assumes that the ac gate voltage
acting on the quantum dot is equal to the ac voltage at the voltage antinode
(i.e., the coupling capacitor) of the microwave resonator. In reality, the ac
gate voltage is likely smaller: the gate finger is attached close but not at
the end of the resonator, and along its length the voltage amplitude can vary
as well. Note that the gate electrode is a 100\,nm wide gold strip, and
accordingly has finite Ohmic resistance as well as a geometry not adapted to
$Z_0=50\,\Omega$. This leads to an underestimation of \ncav.

\begin{figure}[t]
\begin{center}
\includegraphics{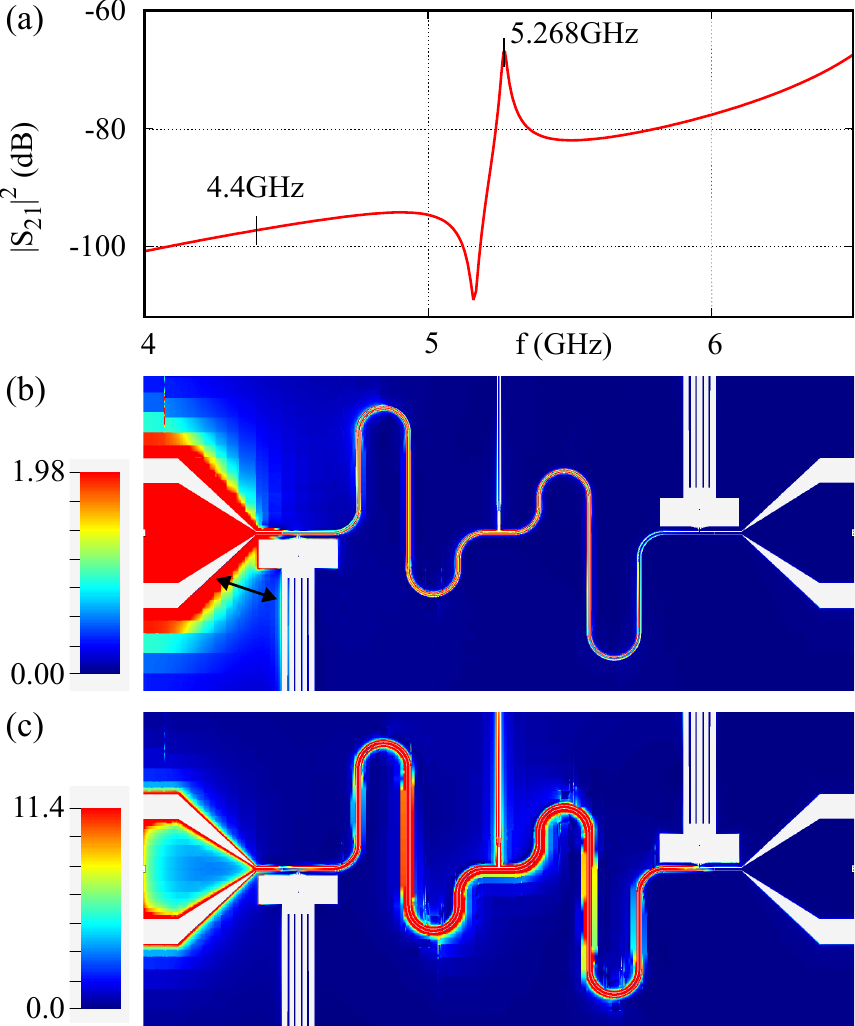}
\end{center}
\caption{Numerical calculation using Sonnet \cite{sonnet} of the GHz response 
of the bare niobium resonator chip geometry (i.e., only patterned niobium on 
the substrate, no further metallizations or structuring). (a) Transmission 
$\left| S_{21}\right|^2$ as function of frequency. The frequencies of (b)
and (c) are marked in the plot. (b), (c) Detail zoom of the rf current
distribution for a drive at (b) $f=4.4\un{GHz}$ (off-resonant) and (c)
$f=5.268\un{GHz}$ (resonant in the calculation) (arbitrary units).
\label{fig:sonnet}\label{figSonnet}
}
\end{figure}
On the other hand, the coplanar waveguide resonator is strongly undercoupled.
This leads to an asymmetry of signal level between input (drive) and
output (detection) port, for both the resonant case and for the off-resonant 
case. Fig.~\ref{figSonnet} illustrates this with a numerical calculation of our 
bare resonator geometry (i.e., only the niobium layer and no further 
fabrication) using Sonnet Professional \cite{sonnet, ieeemtt-rautio-1987, 
book-harrington-1993}. Modeling the full chip is significantly more difficult 
because of the large differences in scale between meander filters and 
electrodes on one hand and the full coplanar waveguide resonator on the other 
hand \cite{stepwisefab}. The signal transmission $\left| S_{21}\right|^2$, 
plotted as function of frequency in Fig.~\ref{figSonnet}(a), reaches in the 
calculation a maximum of $-67\un{dB}$ at resonance $f\simeq 5.27\un{GHz}$.

The current distribution of the Sonnet calculation is plotted in
Fig.~\ref{figSonnet}(b) for $f=4.4\un{GHz}$, i.e., a strongly detuned red 
sideband drive as in our measurement. A closer look shows that in particular 
for such an off-resonant drive signal, direct crosstalk between the input port 
and the closeby nanotube contact electrodes (indicated by a black arrow) can
cause additional ac signals on the nanotube. In the evaluation of
Section~\ref{impact-ghz-dot} this can lead to an overestimation of \ncav.
Future device design will have to take this error mechanism into account.

An alternative estimate for the resonator photon number can be performed using 
the nominal attenuation and amplification values of the microwave setup. 
Following \cite{jap-sage-2011} we calculate the photon number in the resonant 
case $f=\fcav=5.74\un{GHz}$, $Q=497$, for a VNA output power of $25\un{dBm}$.  
With the attenuation in the input cable of $-53\un{dB}$ excluding the cable 
loss, and a cable loss in both input and output each of $-8\un{dB}$, the 
resulting incoming power at the resonator input port is $-36\un{dBm}$, see 
also the circut scheme of Fig.~\ref{figCoplanarRes}(a).

At resonance, we measure on the VNA a transmission attenuation of 
$-49.6\un{dB}$; considering the $90\un{dB}$ amplification of low temperature and 
room temperature amplifiers and again the cable loss, we obtain $-106.6\un{dBm}$ 
as the power leaving the resonator. 

This results in an insertion loss of 
\begin{equation}
 \text{IL} = \frac{P_\text{out}}{P_\text{in}} = -70.6 \un{dB}, 
\end{equation}
allowing us to calculate the circulating power
\begin{equation}
 P_\text{circ} = P_\text{in} Q\cdot 10^{\text{IL}/20}/\pi = 11.7 \un{nW},
\end{equation}
and finally the photon number in the resonator for the resonant driving case
\begin{equation}
 n_\text{res} =\frac{P_\text{circ}}{h \fcav^2}=537 000.
\end{equation}

To obtain the photon number for the case of red sideband drive we multiply 
$n_\text{res}$ with the ratio $\left| S_{21}(f)/S_{21}(\fcav) \right|^2$, 
resulting in $n(25\un{dBm})=72$, or at the drive power of the OMIT experiments
$n(20\un{dBm})=23$, respectively. 

The value of $n_c=21300$ derived from the calibration in the main text is much 
larger. However, it is consistent with the broadening of the Coulomb 
oscillation in \ref{figOscPotential}(a). Additionally, a much smaller photon 
number would result in an even much larger single photon coupling via the OMIT 
experiment than already found, at risk of an equally large overestimation of 
our already large Coulomb-blockade enhancement of the optomechanical coupling.
While this naturally gives rise to optimism, it will have to be confirmed in
future experiments.

\section{Cavity pull parameter}

\label{app-G}
The frequency shift per displacement or cavity pull parameter $G$ is defined as
\begin{equation}
G=\left. \frac{\partial\wcav}{\partial x} \right|_{x=0}
\end{equation}
In a microwave optomechanical system with a deflection-dependent cavity 
(replacement) capacitance \CRLC, using the relation $\wcav=1/\sqrt{\CRLC\LRLC}$ 
it can be written as
\begin{equation}
 G = \frac{\wcav}{2\CRLC} \left. \frac{\partial\CRLC}{\partial x} \right|_{x=0}
\end{equation}
Applying the same logic as in \cite{optomechanics}, Supplement, Eqns. (26-29),
we can translate a capacitance modulation into an effective gate volage
modulation via $\Cg\,\partial\Vg = \Vg\,\partial\Cg$ and write out $G$ as
\begin{equation}
 G = \frac{\wcav}{2\CRLC} \, \frac{\partial \CRLC}{\partial \Vg} \,
 \frac{\Vg}{\Cg} \, \frac{\partial \Cg}{\partial x}
\end{equation}
where \Cg\ is the geometric capacitance between resonator (gate) and nanotube
(quantum dot).

\bibliography{paper}

\end{document}